\documentclass[11pt,a4paper]{article}

\usepackage{amsmath}
\usepackage{amssymb}
\usepackage{graphicx}
\usepackage{bm}

\usepackage{mathrsfs}
\usepackage{cite}

\newcommand*\xbar[1]{%
  \hbox{%
    \vbox{%
      \hrule height 0.5pt 
      \kern0.5ex
      \hbox{%
        \kern-0.1em
        \ensuremath{#1}%
        \kern-0.1em
      }%
    }%
  }%
}

\def\sB{\stackrel{\frown}{\square}}
\def\bsB{\stackrel{\bm{\frown}}{\bm{\square}}}

\numberwithin{equation}{section}

\begin{document}

\title{One-loop finiteness in higher-derivative\\
$6D$, ${\cal N}=(1,0)$ super Yang-Mills -- hypermultiplet system}

\author{
I.L.Buchbinder\footnote{buchbinder@theor.jinr.ru}\ ${}^{a,b}$,
A.S.Budekhina\footnote{budekhina@theor.jinr.ru}\ ${}^{a}$,
E.A.Ivanov\footnote{eivanov@theor.jinr.ru}\ ${}^{a,c}$, and
K.V.Stepanyantz\footnote{stepan@m9com.ru}\
${}^{d}$\\
\\
${}^a${\small{\em Joint Institute for Nuclear Research}}, {\small{\em  Bogoliubov Laboratory of Theoretical Physics}}\\
{\small{\em 141980, Dubna, Moscow region, Russia}}\\
${}^b${\small{\em Center of Theoretical Physics, Tomsk State
Pedagogical University, 634061, Tomsk, Russia}}\\
${}^c${\small{\em Moscow Institute of Physics and Technology, 141700 Dolgoprudny, Moscow region, Russia}}\\
${}^d${\small{\em Moscow State University}}, {\small{\em  Faculty of Physics, Department  of Theoretical Physics}}\\
{\small{\em 119991, Moscow, Russia}}\\
}

\date{}

\maketitle

\begin{abstract}
We employ the harmonic superspace methods to study a six-dimensional
$\mathcal{N}=(1,0)$ supersymmetric gauge theory with higher
derivatives coupled to a hypermultiplet in the adjoint
representation. By introducing a certain non-minimal interaction
between the gauge multiplet and the hypermultiplet, we demonstrate
that the one-loop divergences in gauge superfield sector,
which are present in the conventional formulation, are canceled. The
resulting theory is off-shell one-loop finite in this
sector, while preserving the gauge invariance and
$\mathcal{N}=(1,0)$ supersymmetry. The cancelation mechanism is
explicitly verified using both the background field method and the
supergraph techniques. Thus, we present an example of the
higher-derivative supersymmetric gauge theory in six dimensions
which is finite in the vector multiplet sector.
\end{abstract}

\vspace{0.6cm}

\hspace{3.4cm} \verb"Dedicated to the bright memory of Kellogg S. Stelle"

\vspace{0.6cm}

\allowdisplaybreaks

\section{Introduction}
\hspace*{\parindent}

Supersymmetric field theories in higher dimensions attract a
considerable attention due to their intimate relationships with
string theory and brane dynamics, as well as in connection with the
quest for ultraviolet-complete quantum field theories (see e.g.
\cite{Howe:1983jm,Howe:2002ui,Bossard:2009sy,Bossard:2009mn,Fradkin:1982kf,
Marcus:1983bd,Smilga:2016dpe,Bork:2015zaa}).

The ordinary six-dimensional gauge theories are non-renormalizable by power counting because the gauge coupling constant $g$ is dimensionful, $[g] = -1$ in mass units. This leads to the uncontrollable proliferation of
counterterms at the quantum level. The well-established strategy to improve the ultraviolet behavior is to add some higher-derivative terms to the action, which can render the theory renormalizable or even finite
\cite{Pais:1950za, Stelle:1976gc}. In the supersymmetry context, the higher-derivative extension of $6D$, $\mathcal{N}=(1,0)$ Yang-Mills theory was constructed in harmonic superspace
\cite{Galperin:1984av,Galperin:1985bj,Galperin:1985va,Galperin:2001uw} in Ref.~\cite{Ivanov:2005qf}. The classical action of such a theory reads \footnote{The detailed description of the notations is presented in
Sect.~\ref{Section_Model}. Taking into account their difference, the sign in Eq. (\ref{HD_Theory})
corresponds to the one in Ref. \cite{Ivanov:2005qf}.}

\begin{equation}\label{HD_Theory}
S_{\text{HD}} = -\frac{1}{2e_0^2} \mathrm{tr} \int d\zeta^{(-4)}\, \big( F^{++} \big)^2,
\end{equation}

\noindent where $F^{++}$ is the covariantly analytic superfield
strength of the gauge superfield $V^{++}$, and $e_0$ is a
dimensionless coupling constant. The component form of this action
includes the term $(\nabla^M F_{MN})^2$ in the gauge-field sector,
thus implying the fourth-order kinetic term. The theory proves to be
power-counting renormalizable, and its one-loop divergences have
been computed both in the component and the superfield formulations
\cite{Ivanov:2005qf,Casarin:2019aqw,Buchbinder:2020tnc,Buchbinder:2020ovf}.

Such higher-derivative gauge theory can be generalized  by adding
couplings to the matter hypermultiplet(s). In $6D$,
$\mathcal{N}=(1,0)$ harmonic superspace, hypermultiplets are
described by analytic superfield $q^+$ and its conjugate $\widetilde
q^+$ \cite{Galperin:2001uw}. The minimal coupling to a gauge
multiplet is implemented via the covariant derivative $\nabla^{++}
q^+ = D^{++} q^+ + i V^{++} q^+$, and the classical gauge invariant
action for the hypermultiplet with the standard component kinetic
terms is written as

\begin{equation}
S_{\text{hyper}} = -\int d\zeta^{(-4)} \, \widetilde{q}^+ \nabla^{++} q^+.
\end{equation}

\noindent However, the addition  of the hypermultiplets brings about
a new source of ultraviolet divergences. In the harmonic superspace,
the one-loop quantum corrections for various $6D$ supersymmetric
theories containing the hypermultiplets have been carefully
analyzed, e.g., in
\cite{Buchbinder:2016gmc,Buchbinder:2016url,Buchbinder:2017ozh,Budekhina:2021kbf},
where it was shown that the hypermultiplet sector contributes to the
renormalization of the gauge coupling and may also induce divergent
terms which depend on the hypermultiplet fields themselves. In
particular, in ref. \cite{Budekhina:2021kbf} the one-loop
divergences were computed for a model with an arbitrary $q^+$
hypermultiplet self-interaction,
and non-vanishing divergent contributions for the  on-shell background $q^+$ superfields were found.

A natural question is as to whether a reasonable choice of the
hypermultiplet representation and interaction could improve the
ultraviolet behavior, possibly giving rise even to a finite theory.
For instance, in the four-dimensional case, choosing the
hypermultiplet in the adjoint representation, one obtains the
maximally extended $4D, {\cal N}=4$ supersymmetric Yang--Mills (SYM)
theory, which is finite to all
loops\cite{Sohnius:1981sn,Grisaru:1982zh,Howe:1983sr,Mandelstam:1982cb,Brink:1982pd},
the finiteness being secured through precise cancelations among the
gauge and matter divergent contributions.  In six dimensions,
however, it is hard to find examples of finite theories, partly
because of the stringent constraints imposed by the dimensionality
reasonings and the lack of extra supersymmetry in the presence of
matter. In particular, $6D$, ${\cal N}=(1,1)$ SYM theory (which
amounts to the minimal interaction of the conventional ${\cal
N}=(1,0)$ SYM theory \cite{Zupnik:1986da} with the hypermultiplet in
the adjoint representation) is not renormalizable. However, an
interesting model could come out \cite{Buchbinder:2025aac} when the
hypermultiplet in the adjoint representation interacts with the
higher-derivative gauge theory (\ref{HD_Theory}). Unlike standard
$6D$, ${\cal N}=(1,0)$ theories \cite{Kuzenko:2017xgh}, it is not
anomalous \cite{Smilga:2006ax} and, according to ref.
\cite{Buchbinder:2025aac}, presumably possesses the all-loop exact
$\beta$-function similar to the NSVZ $\beta$-function in ${\cal
N}=1$ supersymmetric gauge theories
\cite{Novikov:1983uc,Jones:1983ip,Novikov:1985rd,Shifman:1986zi}.
Moreover, certain properties of quantum corrections in this case
resemble similar properties of $4D$, ${\cal N}=1$ SYM theories
regularized by the higher-order covariant derivatives
\cite{Slavnov:1971aw,Slavnov:1972sq,Slavnov:1977zf} in the
supersymmetric formulation
\cite{Krivoshchekov:1978xg,West:1985jx,Buchbinder:2015eva} (see
\cite{Stepanyantz:2019lyo,Stepanyantz:2023jot} and references
therein). Unfortunately, the theory considered in
\cite{Buchbinder:2025aac} is not one-loop finite. So, it is of
considerable interest to explore  whether it is possible to modify
this theory so as to make it at least one-loop finite. Note that the
one-loop finite $4D$ supersymmetric theories (see
\cite{Heinemeyer:2019vbc} and references therein) are of special
interest because the one-loop finiteness in this case ensures the
all-loop finiteness under a certain choice of the renormalization
prescription
\cite{Kazakov:1986bs,Ermushev:1986cu,Lucchesi:1987he,Lucchesi:1987ef}
(see also
\cite{Parkes:1984dh,Jones:1983vk,Jones:1984cx,Grisaru:1985tc,Stepanyantz:2021dus}).

The reasons just adduced explain why seeking for  $6D$
one-loop finite supersymmetric theories may constitute a quite
interesting task. Here we propose a modified $6D$,
$\mathcal{N}=(1,0)$ higher-derivative gauge theory with a
hypermultiplet in the adjoint representation that exhibits one-loop finiteness in the gauge multiplet sector. The source of
higher derivatives in this theory, apart from the gauge field action
\eqref{HD_Theory}, is also a modified hypermultiplet action which
involves higher derivatives and the non-minimal interaction of the form

\begin{equation}
S_{\text{int}} = - \frac{2i{\xi}}{e_0^2}\,\mbox{tr} \int d\zeta^{(-4)}\, \tilde{q}^+ \, [F^{++}, \, q^+],
\end{equation}

\noindent where ${\xi}$ is the relevant numerical constant. This interaction term was earlier discussed in \cite{Ivanov:2005kz}. It is gauge-invariant and preserves $\mathcal{N}=(1,0)$
supersymmetry. Remarkably, in this theory, like in $6D$,
$\mathcal{N}=(1,0)$ SYM minimally coupled to the adjoint
hypermultiplet, the one-loop divergences in the gauge multiplet
sector coming from the gauge superfield, hypermultiplet, and ghosts
are canceled among themselves for a certain choice of $\xi$.
To the best of our knowledge, this is the first example of the
higher-derivative gauge theory in six dimensions with the vanishing one-loop $\beta$-function.

Our analysis employs the harmonic superspace formalism
\cite{Galperin:2001uw}, which maintains manifest $\mathcal{N}=(1,0)$
supersymmetry at all steps of quantum calculations. We quantize the
theory by making use of the background superfield method, which
guarantees gauge invariance of the effective action
\cite{Buchbinder:1997ya,Buchbinder:2001wy}. The one-loop effective
action is computed by using the supergraph techniques, and the
relevant divergent parts are extracted through the dimensional
reduction. We verify that for a certain value of $\xi$ the total
divergent contribution vanishes, thus implying the one-loop
finiteness.

The paper is organized as follows. In Sect.~\ref{Section_Model} we
review the harmonic superspace formulation of $\mathcal{N}=(1,0)$
gauge theories and construct the classical action of the
higher-derivative $\mathcal{N}=(1,0)$ gauge theory coupled to a
hypermultiplet in the adjoint representation via the non-minimal
interaction $\sim \mbox{tr}\big(\tilde{q}^+ [F^{++},\, q^+]\big)$.
Sect.~\ref{Background_field_method} is devoted to the background
field quantization of the theory under consideration. As a
first step to renormalization of the theory under consideration, we
are going to study the structure of effective action only in the gauge
multiplet sector. Therefore, we introduce the background field only
for gauge superfield. In Sect.~\ref{Divergences} we compute the
one-loop divergences using two complementary methods: ({\bf i}) the
supergraph techniques (Subsect.~\ref{Supergraph_method}) and ({\bf
ii}) the manifestly covariant proper-time method
(Subsect.~\ref{Manifest_covariant_method}). Both approaches confirm
that for the specific value $\xi=1$ all divergences are
mutually canceled, leading to a theory with the vanishing one-loop $\beta$-function. Finally, in
Conclusions we summarize the results, discuss their significance,
and outline possible extensions of our study. The component form of the action under consideration is discussed in Appendix \ref{Appendix_Components}, while
the technical details of the supergraph calculations are collected in Appendix \ref{Appendix_Supergraphs}.

\section{Harmonic superspace formulation of the higher-derivative \\ $6D,\, \mathcal{N}=(1,0)$ gauge theory with hypermultiplet}
\hspace{\parindent}\label{Section_Model}

Our analysis employs the harmonic superspace formalism, which offers a natural setting for constructing and quantizing
$6D$, $\mathcal{N}=(1,0)$ supersymmetric theories with  preserving manifest supersymmetry and gauge invariance throughout the calculations.

$6D$, $\mathcal{N}=(1,0)$ harmonic superspace was introduced in the works \cite{Howe:1985ar,Zupnik:1986da}.
Following ref. \cite{Bossard:2015dva}, we will parametrize the harmonic superspace by the coordinates $(x^M, \theta^{a}_i, u^{\pm i})$, where $x^M$ ($M=0,\ldots,5$)
are $6D$ Minkowski coordinates  and $\theta^{a}_i$ ($a=1,\ldots,4$; $i=1,2$) are the $\mathcal{N}=(1,0)$ Grassmann coordinates.
The harmonic variables $u^{\pm i}$ are subject to the constraints

\begin{equation}
u^{+i} u_i^- = 1, \qquad u_i^- \equiv (u^{+i})^*.
\end{equation}

\noindent
Using these, one defines the harmonic derivatives

\begin{equation}
D^{++} = u^{+i} \frac{\partial}{\partial u^{-i}}; \qquad D^{--} = u^{-i} \frac{\partial}{\partial u^{+i}}, \qquad D^0 = u^{+i} \frac{\partial}{\partial u^{+i}} - u^{-i} \frac{\partial}{\partial u^{-i}},
\end{equation}

\noindent
which generate $SU(2)$ algebra. A key feature is the existence of an analytic subspace which is closed under the supersymmetry transformations and amounts to the set of coordinates

\begin{equation}
x^M_{{\cal A}} = x^M + \frac{i}{2}\theta^{-}\gamma^M \theta^+, \qquad \theta^{+ a} = u^{+}_i \theta^{ai}, \qquad  u^{\pm i}\,,
\end{equation}

\noindent
where $\gamma^M$ are the $6D$ gamma matrices. It is also useful to introduce harmonic projections of the spinor covariant derivatives,

\begin{equation}
D^+_a = u^{+}_i D_{a}^i, \qquad D^-_a = u^{-}_i D_{a}^i.
\end{equation}

\noindent
Their only non-vanishing anticommutator is $\{D^+_a, D^{-}_b\} = i(\gamma^M)_{ab}\partial_M$. Since the derivatives $D^+_a$ anticommute, any product of four of them
(defined at the same harmonic ``point''  $u$) reduces to the unique expression,

\begin{equation}
(D^+)^4 = -\frac{1}{24}\varepsilon^{abcd} D_a^+ D_b^+ D_c^+ D_d^+.
\end{equation}

We include the integration over harmonic variables $u^{+i}$ into the analytic superspace integration measure, but not into the full harmonic superspace measure,

\begin{equation}
\int d\zeta^{(-4)} \equiv \int d^6x\, d^4\theta^+ du, \qquad \int d^{14}z \equiv \int d^6x\,d^8\theta = \int d^6x\,d^4\theta^{+} (D^+)^4.
\end{equation}

\noindent
For convenience, the integration over harmonic variables in the second case will be performed separately.

In $6D$, ${\cal N}=(1,0)$ harmonic superspace, the pure $6D$, ${\cal N}=(1,0)$ supersymmetric higher-derivative SYM theory is described by the action \cite{Ivanov:2005qf}

\begin{equation}\label{Usual_HD}
S = \pm \frac{1}{2e_0^2} \mbox{tr} \int d\zeta^{(-4)} (F^{++})^2\,,
\end{equation}

\noindent with the dimensionless coupling constant $e_0$. The harmonic gauge superfield strength $F^{++}$ is defined by the expression

\begin{equation}\label{F++}
F^{++} \equiv (D^+)^4 V^{--},
\end{equation}

\noindent where the non-analytic superfield $V^{--}$ is defined as

\begin{equation}\label{V--_Definition}
V^{--}(z,u) \equiv \sum\limits_{n=1}^\infty (-i)^{n+1} \int du_1 du_2 \ldots du_n \frac{V^{++}(z,u_1) V^{++}(z,u_2)
\ldots V^{++}(z,u_n)}{(u^+u_1^+)(u_1^+ u_2^+) \ldots (u_n^+ u^+)}.
\end{equation}

\noindent The gauge superfield $V^{++}(z,u) = V^{++A} t^A$ obeys the analyticity condition

\begin{equation}
D^+_a V^{++} = 0,
\end{equation}

\noindent and is real with respect to a generalized tilde-conjugation, $\widetilde{V^{++}} = V^{++}$.
Note that the non-analytic prepotential $V^{--}$ and the superfield strength $F^{++}$ are also real, while $F^{++}$ is analytic by its definition.
Moreover, it satisfies the off-shell condition

\begin{equation}\label{Equation_For_F}
\nabla^{++} F^{++} \equiv D^{++} F^{++} + i [V^{++}, F^{++}] = 0\,,
\end{equation}

\noindent
which is a consequence of both the analyticity and the harmonic zero-curvature condition \cite{Galperin:2001uw},

\begin{equation}\label{Flatness}
D^{++} V^{--} - D^{--}V^{++} + i [V^{++}, V^{--}] = 0\,.
\end{equation}

The component structure of the action (\ref{Usual_HD}) was discussed in \cite{Ivanov:2005qf}.
The $6D$ gauge multiplet $V^{++}$ in the Wess-Zumino gauge involves the vector field $A_M$, the  Weyl fermion $\psi^{a i }$ and an $SU(2)$
triplet of scalar fields $d^{(ij)}$. In the component notation the action (\ref{Usual_HD})
contains four derivatives in the kinetic term of the gauge field $A_M$ and three derivatives in the kinetic term of the gaugino $\psi^{ai}$.
The scalar fields $d^{ij}$ are also dynamical, with the standard two derivatives in the kinetic term.

In this paper we use the Hermitian gauge group generators $t^A$ which are normalized by the condition $\mbox{tr}(t^A t^B) = \delta^{AB}/2$.
The gauge transformations in the harmonic superspace are parametrized by a real (with respect to the tilde-conjugation) analytic superfield $\lambda = \lambda^A t^A$,
$\widetilde{\lambda} = \lambda$,
and are written in the form

\begin{eqnarray}\label{Gauge_Invariance}
V^{\pm\pm} \to e^{i\lambda} V^{\pm\pm} e^{-i\lambda} - i e^{i\lambda} D^{\pm\pm} e^{-i\lambda},\qquad\quad
F^{++} \to e^{i\lambda} F^{++} e^{-i\lambda}.\qquad
\end{eqnarray}

We will consider the $6D$, ${\cal N}=(1,0)$ supergauge higher-derivative theory interacting with the higher-derivative hypermultiplet described
by the analytic superfield $q^+ = q^{+A} t^A$. For the latter  we assume the action which, besides the ``minimal'' part (coinciding with the expression $S_2$ in Ref. \cite{Ivanov:2005kz}) characterized by the same coupling constant $e_0$ as the gauge part $\sim (F^{++})^2$, contains the interaction term proportional to $F^{++}$ (coinciding with the expression $S_1$ in Ref. \cite{Ivanov:2005kz})

\begin{equation}\label{Action}
S = \pm \frac{1}{2e_0^2} \mbox{tr} \int d\zeta^{(-4)} (F^{++})^2 - \frac{2}{e_0^2}\,\mbox{tr} \int d\zeta^{(-4)} \widetilde{q}^+ \left(\sB \nabla^{++} q^+ +i\xi [F^{++},\, q^+]\right),
\end{equation}

\noindent with an arbitrary dimensionless numerical parameter $\xi$. The covariant analytic d'Alembertian $\sB$ in six dimensions is defined as

\begin{equation}\label{Box}
\sB\, \, \equiv \, \frac{1}{2}(D^+)^4(\nabla^{--})^2,
\end{equation}

\noindent
where $\nabla^{\pm\pm} =D^{\pm\pm} + iV^{\pm\pm}$. When acting on the analytic superfields, this operator is reduced to \cite{Buchbinder:2014sna}:

\begin{eqnarray}
\label{Box_2} \sB \,\, = \Box + iW^{+a} \nabla^{-}_a + iF^{++} \nabla^{--} - \frac{i}{2}(\nabla^{--} F^{++}).
\end{eqnarray}

\noindent
Here the covariant derivatives are defined by the equations

\begin{equation}
[\nabla^{--},D_a^+] = \nabla_a^-;\qquad \{D_a^+,\nabla_b^-\} = i(\gamma^M)_{ab}\nabla_M,
\end{equation}

\noindent
$\Box=\eta^{MN}\nabla_M\nabla_N$, and $W^{+a}$ is the spinorial covariant superfield strength defined as

\begin{equation}
W^{+a}=-\frac{1}{6}\varepsilon^{abcd}D^+_b D^+_c D^+_d V^{--}\,.
\end{equation}

The component form of the action (\ref{Action}) is discussed in Appendix \ref{Appendix_Components}.

As the basic result of this work, we will demonstrate that for the specific value of the parameter $\xi$ the theory becomes finite at the one-loop level in the gauge multiplet sector.
To establish this property and to understand how the underlying cancelation of divergences comes about, we proceed to computing
and analyzing the one-loop divergences for an arbitrary value of $\xi$.

\section{Background field quantization}
\label{Background_field_method}

\hspace{\parindent} The harmonic superspace technique ensures the
manifestly ${\cal N}=(1,0)$ supersymmetric quantization procedure.
It is also convenient to use the background superfield method
\cite{DeWitt:1965jb,DeWitt:1967ub} (see also
\cite{Kallosh:1974yh,Arefeva:1974jv,Abbott:1980hw,Abbott:1981ke}),
which provides a manifestly gauge invariant effective action. In
$6D$, ${\cal N}=(1,0)$ harmonic superspace it is formulated
similarly to the $4D$, ${\cal N}=2$ case treated in
\cite{Buchbinder:1997ya,Buchbinder:2001wy}. In particular, the
background-quantum splitting of the gauge superfield is linear,

\begin{equation}\label{Splitting}
V^{++} = \bm{V}^{++} + v^{++},
\end{equation}

\noindent where $\bm{V}^{++}$ is the background gauge superfield and
$v^{++}$ is the quantum gauge superfield\footnote{Such a
splitting is enough to study the divergences in the gauge superfield
sector.}. After making the substitution (\ref{Splitting}) in the
action (\ref{Action}), the gauge invariance (\ref{Gauge_Invariance})
amounts to the two types of transformations. The background gauge
invariance,

\begin{equation}\label{Background_Symmetry}
\bm{V}^{++} \to e^{i\lambda} \bm{V}^{++} e^{-i\lambda} - i e^{i\lambda} D^{++} e^{-i\lambda}, \qquad v^{++} \to e^{i\lambda} v^{++} e^{-i\lambda}\,
\end{equation}

\noindent
remains a manifest symmetry of the effective action if the gauge-fixing term is chosen to be invariant under (\ref{Background_Symmetry}), while the quantum gauge invariance,

\begin{equation}
\bm{V}^{++} \to \bm{V}^{++}, \qquad v^{++} \to e^{i\lambda} (v^{++} + \bm{V}^{++}) e^{-i\lambda} -\bm{V}^{++} - i e^{i\lambda} D^{++} e^{-i\lambda}\,,
\end{equation}

\noindent
is broken by the gauge-fixing procedure down to the invariance under the BRST transformations.

It is convenient to define the background covariant derivatives

\begin{equation}\label{Background_Covariant_Derivatives}
\bm{\nabla}^{++} = D^{++} + i \bm{V}^{++}, \qquad \bm{\nabla}^{--} = D^{--} + i \bm{V}^{--}
\end{equation}

\noindent
and introduce the notations

\begin{equation}\label{BBox}
\bsB\, \, \equiv \, \frac{1}{2}(D^+)^4(\bm{\nabla}^{--})^2,\qquad \bm{F}^{++} \equiv (D^+)^4\bm{V}^{--},
\end{equation}

\noindent
where $\bm{V}^{--}$ is constructed exactly as in Eq. (\ref{V--_Definition}), but with the background gauge superfield in the right-hand side.
Evidently, if the background covariant derivatives act on a superfield in the adjoint representation (e.g., on $v^{++}$), then the gauge superfields
should be expanded over the generators of the adjoint representation:

\begin{equation}
\bm{\nabla}^{\pm\pm} v^{++} = D^{\pm\pm} v^{++} + i [\bm{V}^{\pm\pm}, v^{++}].
\end{equation}

The generating functional (\ref{Action}) can be written in the form

\begin{equation}
Z[\,\mbox{Sources}, \bm{V}^{++}] = \int Dv^{++}Dq^{+} Db\, Dc\, D\varphi\, D\chi^{(+4)} D\sigma\, \exp\Big\{i\Big(S + S_{\mbox{\scriptsize sources}}+\cdots\Big)\Big\},
\end{equation}

\noindent
where dots stand for the rest of $S_{\mbox{\scriptsize total}}$. The source term is defined as

\begin{equation}
S_{\mbox{\scriptsize sources}} = \int d\zeta^{(-4)} \Big(J^{++A} v^{++A} + j^{(+3)A}q^{+A} + \widetilde{j}^{(+3)A}\widetilde{q}^{+A} + \ldots\Big),
\end{equation}

\noindent
where dots denote terms corresponding to various ghost superfields. The effective action $\Gamma[\,\mbox{Fields}, \bm{V}^{++}]$
is defined as the Legendre transform of the generating functional for the connected Green functions $W\equiv -i\ln Z$. Setting all quantum fields
equal to zero, we obtain the manifestly gauge invariant action $\Gamma[\bm{V}^{++}]\equiv \Gamma[\,\mbox{Fields}\to 0, \bm{V}^{++}]$.

\section{Calculation of divergences}
\hspace{\parindent}\label{Divergences}

Divergences for the pure higher-derivative theory (\ref{HD_Theory}) have been calculated in \cite{Ivanov:2005qf,Casarin:2019aqw,Buchbinder:2020tnc,Buchbinder:2020ovf}.
They are given by the expression

\begin{equation}\label{result_pureHD}
\Delta\Gamma_\infty^{(1)} = -\frac{11}{3} \frac{C_2}{\varepsilon(4\pi)^3} \mbox{tr} \int d\zeta^{(-4)}\, (\bm{F}^{++})^2,
\end{equation}

\noindent
where it is assumed that the regularization is made by dimensional reduction \cite{Siegel:1979wq} and $\varepsilon \equiv 6-D$. The constant $C_2$ is defined by the equation $f^{ACD} f^{BCD} = C_2 \delta^{AB}$,
where the structure constants $f^{ABC}$ are given by the commutator of generators, $[t^A, t^B] = i f^{ABC} t^C$.

In particular, the result (\ref{result_pureHD}) includes the ghost contribution which is composed of Faddeev--Popov and Nielsen--Kallosh parts and is equal to

\begin{equation}\label{Ghost_Contribution}
(\Delta\Gamma^{(1)}_\infty)_{\mbox{\scriptsize ghost}} = \frac{C_2}{3\varepsilon (4\pi)^3}\, \mbox{tr} \int d\zeta^{(-4)}\, (\bm{F}^{++})^2.
\end{equation}

\noindent
For the theory containing the hypermultiplet in the adjoint representation without higher derivatives (considered in \cite{Buchbinder:2025aac})
the ghost contribution exactly cancels that of the hypermultiplet. Then the one-loop divergences take the form

\begin{equation}\label{Divergent_HD}
(\Delta\Gamma^{(1)}_\infty)_{\mbox{\scriptsize HD+hyper}} = (\Delta\Gamma^{(1)}_\infty)_{\mbox{\scriptsize gauge}} = - \frac{4C_2}{\varepsilon (4\pi)^3} \mbox{tr} \int d\zeta^{(-4)}\, (\bm{F}^{++})^2
\end{equation}

\noindent
and do not vanish. In fact, this expression is a divergent contribution produced by the one-loop superdiagrams with a loop of the quantum gauge superfield.

In this paper we calculate the hypermultiplet contribution for the
action (\ref{Action}) and ask whether it is possible to find such a
value of ${\xi}$ for which the divergences in the gauge multiplet sector vanish. The answer proves
to be in affirmative.

Like in the conventional  $6D, \, {\cal N}=(1,0)$ SYM theory
\cite{Buchbinder:2016url, Buchbinder:2017ozh}, there are two ways to
calculate divergent terms for this action. One of them is based on
the superfield proper-time technique and preserves manifest gauge
invariance at all steps of the calculation. Another amounts to the
direct calculation of the relevant Feynman supergraphs with invoking
gauge invariance at the final stage. We first compute the one-loop
divergences using the supergraph technique. The result is then
cross-checked via the manifestly covariant background field method.

\subsection{Supergraph method of calculation}
\hspace{\parindent}\label{Supergraph_method}

\begin{figure}[h]
\begin{picture}(0,3.1)
\put(2.4,0.2){\includegraphics[scale=0.22]{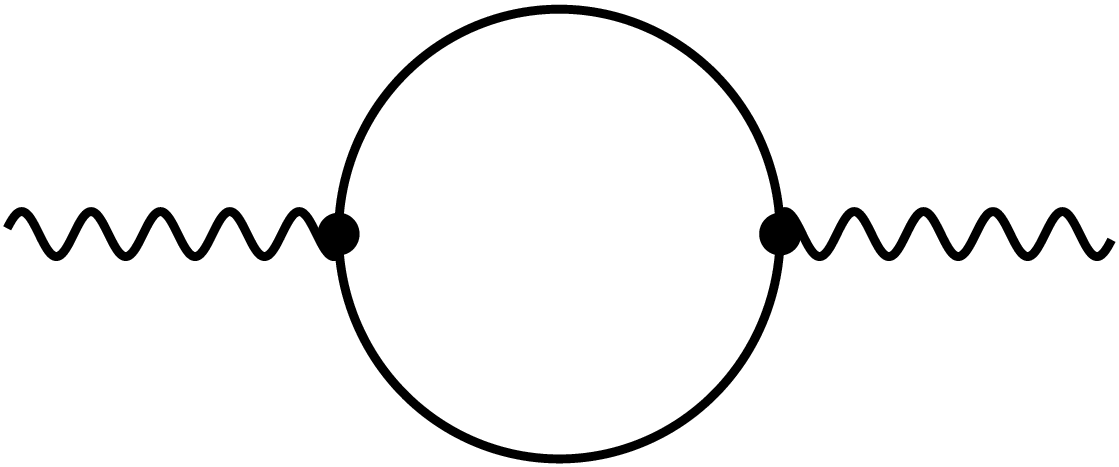}}
\put(2,1.8){(1)}
\put(9.9,-0.2){\includegraphics[scale=0.22]{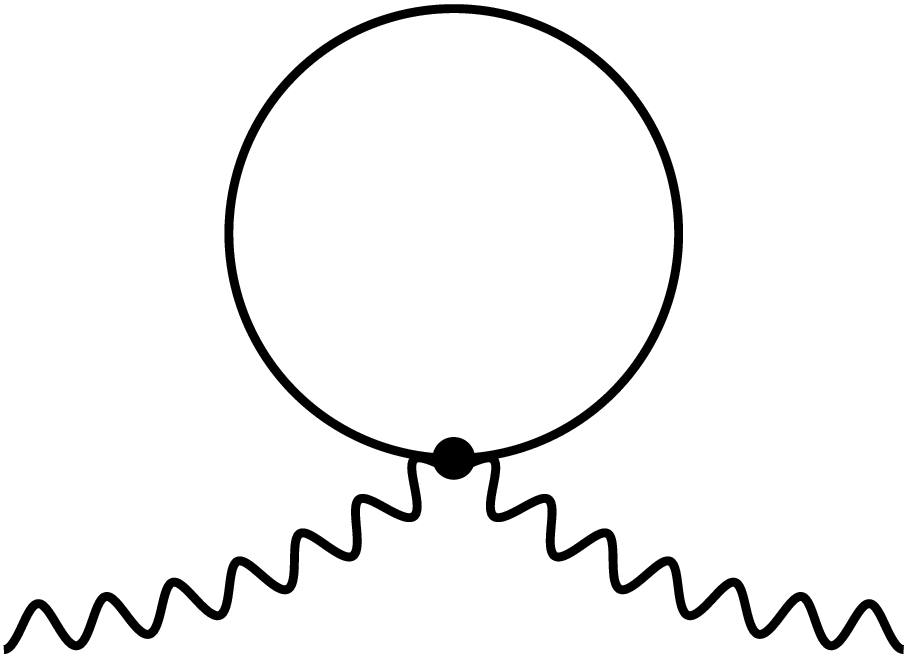}}
\put(9.5,1.8){(2)}
\end{picture}
\caption{Supergraphs with a hypermultiplet loop contributing to the divergent part of the one-loop effective action.}\label{Figure_Diagrams}
\end{figure}

We will calculate only the one-loop divergences produced by the supergraphs containing a loop of the hypermultiplet, because all other contributions
to the divergent part of the effective action are the same as in the theory considered in ref. \cite{Buchbinder:2025aac}. The one-loop contribution
of the hypermultiplet to the two-point Green function of the background gauge superfield for the theory (\ref{Action}) is given by the supergraphs depicted on Fig. \ref{Figure_Diagrams}.
The external wavy lines in these supergraphs correspond to the background gauge superfield $\bm{V}^{++}$, while the solid lines denote the propagators of the hypermultiplet $q^{+}$,

\begin{equation}
- \frac{\delta^2 Z_0}{\delta j_1^{(+3)A} \delta \widetilde j_2^{(+3)B}}\bigg|_{j^{(+3)},\,\widetilde{j}^{(+3)}=0} =  - \frac{ie_0^2}{(u^+_1u^+_2)^3}\, \delta^{AB}\,\frac{1}{(\partial^2)^2} (D_1^+)^4 (D_2^+)^4 \delta^{14}(z_1-z_2),
\end{equation}

\noindent
where $\partial^2 = \eta^{MN}\partial_M \partial_N$,

\begin{equation}
\delta^{14}(z_1-z_2) \equiv \delta^6(x_1-x_2)\, \delta^8(\theta_1-\theta_2),
\end{equation}

\noindent and $Z_0$ is the generating functional of the free
theory.

The hypermultiplet contribution to the effective action is completely determined by that part of the total action which
is quadratic in the superfields $q^+$ and $\widetilde q^+$ and does not contain the quantum gauge superfield $v^{++}$,

\begin{eqnarray}\label{Second_Variation}
S^{(2)} = - \frac{2}{e_0^2} \mbox{tr} \int d\zeta^{(-4)}\,\widetilde{q}^+ \left(\bsB\bm{\nabla}^{++} q^+ + i{\xi}[\bm{F}^{++},\,q^+]\right).
\end{eqnarray}

\noindent Starting from this expression we calculate the one-loop
quantum corrections corresponding to the supergraphs presented in
Fig.~\ref{Figure_Diagrams}. Namely,  from the expression
(\ref{Second_Variation}) we find the expressions for vertices in the
supergraphs under consideration, construct the expressions for them
using the Feynman rules in the harmonic superspace, and obtain the
corresponding contribution to the divergent part of the two-point
Green function of the background gauge superfield. Finally,
resorting to the background gauge invariance of the effective
action, we restore the general result for one-loop divergences
containing all powers of the background gauge superfield. The
details of this calculation are presented in Appendix
\ref{Appendix_Supergraphs}. The final result obtained with the use of the
regularization by dimensional reduction, is given by the expression

\begin{equation}\label{Divergent_Contribution}
\Delta\Gamma^{(1)}_{\infty} = \xi\cdot\frac{4C_2}{\varepsilon (4\pi)^3}\,
\mbox{tr} \int d\zeta^{(-4)}\, (\bm{F}^{++})^2 - \frac{ C_2}{3\varepsilon (4\pi)^3}\, \mbox{tr} \int d\zeta^{(-4)}\, (\bm{F}^{++})^2.
\end{equation}

\noindent The second contribution in Eq.
(\ref{Divergent_Contribution}) exactly cancels the ghost
contribution (\ref{Ghost_Contribution}). Note that this cancelation
occurs in fact already at the level of loop integrals. It is
important because this term and the ghost contributions contain
quadratic divergences which turn out  to cancel each other.
Therefore, for the theory under consideration all divergences
in the gauge superfield sector (at least, in the one-loop
approximation) are logarithmical.

The first contribution in Eq. (\ref{Divergent_Contribution}) cancels
out the remaining divergent contributions (\ref{Divergent_HD})
(coming from the superdiagrams containing a loop of the quantum
gauge superfield),  provided that $\xi=1$. In this case, just due to
the presence of  additional non-minimal interaction term, the theory
becomes off-shell one-loop finite in the gauge superfield
sector.

\subsection{Manifestly covariant method}
\hspace{\parindent}\label{Manifest_covariant_method}

The result obtained in the previous section can be recovered by a shorter calculation based on
a manifestly covariant method of computing divergences (see, e.g., \cite{Buchbinder:2020ovf}).

As the starting point, we rewrite the one-loop contribution under consideration to the effective
action in the form of the functional integral

\begin{equation}
\exp\left(i\Delta\Gamma^{(1)}[\bm{V}^{++}]\right) = \int \mathcal{D}q^+\mathcal{D}\widetilde{q}^+
\exp\left(iS^{(2)}[q^+,\widetilde{q}^+, \bm{V}^{++}]\right), \quad
\end{equation}

\noindent where $S^{(2)}$ is given by the expression (\ref{Second_Variation}). Then the one-loop quantum correction to the effective action is found to be

\begin{equation}\label{Eff_Action1}
\Delta\Gamma^{(1)} = i\,\mathrm{Tr} \ln \big(\bsB \bm{\nabla^{++}}\cdot \mathbf{1} +i {\xi} \bm{F}_{Adj}^{++}\big).
\end{equation}

\noindent
The subscript $Adj$ indicates that the corresponding superfield should be expanded as $\bm{F}_{Adj}^{++} = \bm{F}^{++A} T_{Adj}^A$,
where $(T_{Adj}^A)_{BC} = -if^{ABC}$ are the generators of the adjoint representation.
However, Eq. (\ref{Eff_Action1}) is a formal expression since the argument of the logarithm has a nonvanishing $U(1)$ charge.
In order to find out how to define it  more accurately, let us consider a simpler expression

\begin{equation}\label{Usual_Hypermultiplet}
i\, \mbox{Tr} \ln \bm{\nabla}^{++} = i\, \mbox{Tr} \ln \left(D^{++}+i\bm{V}^{++}\right),
\end{equation}

\noindent
which is the contribution of the conventional hypermultiplet (with the action containing no higher derivatives), and introduce the Green function $G_0^{(1,1)}$ as a solution of the equation

\begin{equation}\label{G0_Definition}
D_1^{++} G_0^{(1,1)}(1|2) = \delta_{\cal A}^{(3,1)}(1|2),
\end{equation}

\noindent
where the analytic delta-function is defined as \cite{Galperin:2001uw}

\begin{equation}
\delta_{\cal{A}}^{(q,4-q)}(1|2)=(D^{+}_2)^4 \delta^{14}(z_1-z_2)\delta^{(q,-q)}(u_1,u_2).
\end{equation}

\noindent
Then the expression (\ref{Usual_Hypermultiplet}) is understood as

\begin{equation}
i\,\mbox{Tr}_{(3,1)} \ln\left(\delta_{\cal A}^{(3,1)} + i\bm{V}_1^{++} G_0^{(1,1)}\right),
\end{equation}

\noindent
where the subscript $1$ points that the superfield $\bm{V}^{++}$ is taken in the harmonic ``point'' $u_1$. The functional trace in this expression includes both the matrix trace and the integration over harmonic superspace,

\begin{equation}
\mathrm{Tr}_{(q,4-q)}\mathcal{O} = \mathrm{tr}\int d\zeta_1^{(-4)}d\zeta_2^{(-4)}\delta_{\cal A}^{(q,4-q)}(2|1)\mathcal{O}^{(q,4-q)}(1|2).
\end{equation}

\noindent
Here, $\mathcal{O}^{(q,4-q)}$ is the kernel of the operator $\mathcal{O}$ acting in the space of analytic superfields with the harmonic $U(1)$ charge $q$.

Similarly, the expression (\ref{Eff_Action1}) can be equivalently represented in the form

\begin{equation}\label{Eff_Action2}
\Delta\Gamma^{(1)} = i\,\mathrm{Tr} \ln\bm{\nabla}^{++}+ i\,\mathrm{Tr} \ln\bsB + i\,\mathrm{Tr}_{(3,1)}\ln \left( \delta_{\cal A}^{(3,1)} + i{\xi}\, {\bsB}{}_1^{-1}\bm{F}_{1, Adj}^{++}G^{(1,1)}\right),
\end{equation}

\noindent
where the Green function of the hypermultiplet $G^{(1,1)}$ satisfies the equation

\begin{equation}
\bm{\nabla}_1^{++}G^{(1,1)}(1|2)= \delta_{\cal{A}}^{(3,1)}(1|2)
\end{equation}

\noindent
and (unlike $G_0^{(1,1)}$ defined by Eq. (\ref{G0_Definition})) depends on the background gauge superfield $\bm{V}^{++}$. The first term in Eq. (\ref{Eff_Action2}) coincides with the contribution of the conventional hypermultiplet (\ref{Usual_Hypermultiplet}). The second term does not produce any divergences, see, e.g., \cite{Buchbinder:2017ozh}. Therefore, it remains to calculate the contribution of the third term. The Green function entering this term is given by the expression

\begin{equation}
G^{(1,1)}(1|2)=\, \bsB{}_1^{-1} (D_1^{+})^4(D_2^{+})^4\, \frac{e^{i\bm{b}_1} e^{-i\bm{b}_2}}{(u_1^+u_2^+)^3}\, \delta^{14}(z_1-z_2),
\end{equation}

\noindent
where $\bm{b}$ is the background bridge superfield defined by the equation $\bm{V}^{++} = - ie^{i\bm{b}} (D^{++} e^{-i\bm{b}})$. Then the third term in Eq. (\ref{Eff_Action2}) can be  represented as a series,

\begin{equation}
i\,\mathrm{Tr}_{(3,1)}\ln \left(\delta_{\cal A}^{(3,1)} + i\xi\bsB{}^{-1}\bm{F}_{Adj}^{++}G^{(1,1)}\right) = i\,\mathrm{Tr}_{(3,1)}\sum_{n = 1}^{\infty}\frac{(-1)^{n-1}}{n} \left(i\xi \bsB{}^{-1} \bm{F}_{Adj}^{++} G^{(1,1)}\right)^{n}.
\end{equation}

\noindent
The only divergent contribution in this expression comes from the term linear in $\xi$, corresponding just to $n=1$ in the sum above,

\begin{eqnarray}
&& -\xi\int d^{(-4)}\zeta_1 \, d^{(-4)}\zeta_2 \, \delta_{\cal{A}}^{(3,1)}(2|1)\,\big(\bsB{}_1^{-1}\big)^{AB}\,(\bm{F}_{Adj,1}^{++})^{BC}\big(\bsB{}_1^{-1}\big)^{CA}\,\nonumber\\
&&\qquad\qquad\qquad\qquad\qquad\qquad\qquad\qquad\qquad \times (D_1^{+})^4(D_2^{+})^4\, \frac{e^{i\bm{b}_1} e^{-i\bm{b}_2}}{(u_1^+u_2^+)^3}\, \delta^{14}(z_1-z_2),\qquad
\end{eqnarray}

\noindent
where we have explicitly written down the gauge group indices. Next, it is necessary to use the expansion for the inverse d'Alembert operator

\begin{equation}
\bsB{}^{-1} = \Box^{-1} - i\, \Box^{-1} \bm{F}_{Adj}^{++} D^{--} \Box^{-1} +\ldots,
\end{equation}

\noindent
where dots stand for those remaining terms in the series which do not contribute to the divergent part of the effective action.
As the next step, the Grassmannian $\delta$-function is removed with the help of the identity

\begin{equation}\label{formula}
\delta^8(\theta_1-\theta_2)\, (D^+_1)^4 (D^+_2)^4 \delta^8(\theta_1-\theta_2) = (u_1^+ u_2^+)^4 \delta^8(\theta_1-\theta_2).
\end{equation}

\noindent
Further, after some algebra involving the well-known identities $(u^+ u^+)=0$ and $(u^+ u^-) = - (u^- u^+)=1$,
the divergent contribution of the third term in Eq. (\ref{Eff_Action2})
to the one-loop effective action can be written in the manifestly covariant form as

\begin{equation}\label{Gauge_Contribution}
\Delta\Gamma^{(1)}_\infty = {\xi} \cdot\frac{4C_2}{\varepsilon (4\pi)^3}\, \mbox{tr} \int d\zeta^{(-4)}\, (\bm{F}^{++})^2.
\end{equation}

\noindent
The first term in Eq. (\ref{Eff_Action2}) produces the divergences

\begin{equation}
- \frac{ C_2}{3\varepsilon (4\pi)^3}\, \mbox{tr} \int d\zeta^{(-4)}\, (\bm{F}^{++})^2\,,
\end{equation}

\noindent which coincide with the contribution of the hypermultiplet
in the case when higher derivatives are absent in its action. As we
have already mentioned, the second term in Eq. (\ref{Eff_Action1})
does not produce divergences. Therefore, the net result for the
divergent contribution to the effective action generated by the
hypermultiplet in the one-loop approximation exactly coincides with
Eq. (\ref{Divergent_Contribution}) derived earlier by a different
techniques.

Thus, the result had been cross-checked by two independent methods and so is correct.

\section{Conclusions}
\hspace*{\parindent}

In this paper, using the harmonic superspace formalism, we have
proposed and investigated a certain $6D,\, \mathcal{N}=(1,0)$
supersymmetric higher-derivative gauge theory with the
hypermultiplet in the adjoint representation. The theory is
specified by the action (\ref{Action}) which, apart from the
standard higher-derivative super-Yang-Mills term, contains higher
derivatives in the hypermultiplet sector and a non-minimal
gauge-invariant interaction between the hypermultiplet and the gauge
superfield strength $F^{++}$, with ${\xi}$ as a parameter of this
non-minimal coupling.

The main result of our work is the demonstration of the fact that,
for the special value ${\xi} = 1$, all off-shell one-loop
ultraviolet divergences in the purely gauge superfield  sector
are exactly canceled among themselves, rendering the theory one-loop
finite. The cancelation just mentioned is between the contributions
coming from the pure higher-derivative gauge sector, the
hypermultiplet sector, and the ghost sector. We have verified this
cancelation by two independent methods: the supergraph technique and
the manifestly covariant background-field formalism. Both approaches
lead to the same expression for the divergences produced by a
hypermultiplet loop, which is given by
Eq.~(\ref{Divergent_Contribution}). At ${\xi}=1$ the non-minimal
coupling term precisely cancels the remaining divergences, yielding
a finite result. Note that the one-loop quadratic divergences in the
theory under consideration cancel each other for any value of
${\xi}$.

To the best of our knowledge, this is the first example of a
one-loop finite higher-derivative gauge theory in six dimensions.
The construction relies crucially on the harmonic superspace
formulation, which preserves manifest $\mathcal{N}=(1,0)$
supersymmetry and gauge invariance at all stages of calculation. The
non-minimal interaction of the very simple form, $\tilde{q}^+
F_{Adj}^{++} q^+$, plays the crucial role in tuning the theory to a
critical point where the quantum cancelations take place.

There remain several open questions and possible extensions, which
deserve a further study. First, it is interesting to investigate the
conditions under which the manifest $\mathcal{N}=(1,0)$
supersymmetry of the higher-derivative theory in $6D$ harmonic
superspace can be enhanced by a hidden $ \mathcal{N}=(0,1) $
supersymmetry to the full $ \mathcal{N}=(1,1) $ supersymmetric
structure\footnote{It should be the appropriate generalization of
the hidden $ \mathcal{N}=(0,1) $ supersymmetry transformations
mixing the on-shell $V^{++}$ and $q^+$ superfields with the
conventional kinetic terms, see \cite{Bossard:2015dva} and
\cite{Buyucli:2021csy}.}. We believe  that this enhancement is
possible only at $\xi=1$. Second, to answer the above
question it would be extremely important to calculate the one-loop
divergences in the purely hypermultiplet and the mixed gauge multiplet -
hypermultiplet sectors and to find out whether no one-loop
divergences at $\xi=1$ are present in these sectors. Third, it would
be tempting to investigate whether the finiteness persists at higher
loops. Although the one-loop cancelation occurs in a  non-trivial
way, the higher-loop divergences might appear unless some additional
symmetries (e.g., extended $ \mathcal{N}=(1,1) $ supersymmetry) are
assumed.

In summary, we have presented an example of the off-shell
six-dimensional higher-derivative supersymmetric gauge theory which is one-loop finite  in the gauge superfield sector. The result
illustrates how the carefully chosen non-minimal couplings can
improve the quantum behavior of the higher-dimensional theories,
offering a new avenue for constructing well-defined quantum field
theories beyond four dimensions.

\appendix

\section*{Appendix}

\section{Component form of the action}\label{Appendix_Components}
\hspace*{\parindent}

In the Wess--Zumino gauge the gauge superfield $V^{++}$ takes the form (see, e.g. ,\cite{Ivanov:2005qf})

\begin{equation}\label{V++_Components}
V^{++} = \frac{1}{2}\theta^{+a}\theta^{+b} (\gamma^M)_{ab} A_M(x_{\cal A}) + 2\sqrt{2} (\theta^+)^3_a \psi^{-a}(x_{\cal A}) +3i (\theta^+)^4 d^{--}(x_{\cal A}),
\end{equation}

\noindent where

\begin{equation}
(\theta^+)^3_a\equiv \frac{1}{6}\varepsilon_{bcda}\theta^{+b}\theta^{+c}\theta^{+d},\qquad (\theta^+)^4\equiv -\frac{1}{24} \varepsilon_{abcd}\theta^{+a}\theta^{+b}\theta^{+c}\theta^{+d},
\end{equation}
and $\psi^{-a} = \psi^{ia}(x_{\cal A}) u^-_i, \quad d^{--}=d^{ik}(x_{\cal A}) u^-_iu^-_k $.

The component expansion of $F^{++}$ was found in \cite{Ivanov:2005qf}. In our notation it reads

\begin{eqnarray}\label{F++_Components}
&&\hspace*{-5mm} F^{++} = i d^{++} + i\sqrt{2} \theta^+ \gamma^M {\cal D}_M \psi^+ + \frac{1}{2} \theta^+ \gamma^M\theta^+ \Big({\cal D}^N F_{NM} + {\cal D}_M d^{+-} + i (\gamma_M)_{ab} \{\psi^{-a}, \psi^{+b}\}\Big) \nonumber\\
&&\hspace*{-5mm}\qquad\quad\ + (\theta^+)_a^3 \sqrt{2} \Big({\cal D}^M {\cal D}_M \psi^{-a} + [d^{--},\psi^{+a}] - 2 [d^{+-},\psi^{-a}] - \frac{1}{2} (\widetilde\gamma^M\gamma^N)^a{}_b [F_{MN},\psi^{-b}]\Big)\nonumber\\
&&\hspace*{-5mm}\qquad\quad\ + (\theta^+)^4 \Big(i{\cal D}_M {\cal D}^M d^{--} +3i[d^{--},d^{+-}] +2 (\gamma^M)_{ab}\{\psi^{-a},{\cal D}_M\psi^{-b}\}\Big).
\end{eqnarray}

\noindent
Here the covariant derivatives and gauge field strength are defined as

\begin{equation}
{\cal D}_M\varphi = \partial_M\varphi + [A_M, \varphi],\qquad F_{MN} = \partial_M A_N - \partial_N A_M + [A_M,A_N].
\end{equation}

\noindent
The expression (\ref{F++_Components}) allows for calculating the component form of the gauge part of the action (\ref{Action}).
Taking into account some differences in our notation compared to \cite{Ivanov:2005qf}, we  for completeness present the corresponding expression,

\begin{eqnarray}
&& S_{\mbox{\scriptsize gauge}} =\pm \frac{1}{2e_0^2} \mbox{tr}\int d\zeta^{(-4)} (F^{++})^2 \to \pm \frac{1}{e_0^2}\mbox{tr} \int d^6x\,\bigg\{({\cal D}^M F_{MN})^2 + i\psi^j \gamma^M {\cal D}_M {\cal D}^N {\cal D}_N \psi_j
\qquad\nonumber\\
&&\qquad\qquad - \frac{1}{2} ({\cal D}_M d_{jk})^2 +i d_{lk} d^{kj} d_j{}^l - 2 d_{jk}\Big(\psi^j\gamma^M{\cal D}_M \psi^k - {\cal D}_M\psi^j \gamma^M \psi^k\Big) + (\psi^j \gamma_M \psi_j)^2\nonumber\\
&&\qquad\qquad + \frac{1}{2} {\cal D}_M \psi^j \gamma^M \widetilde\gamma^K \gamma^L [F_{KL},\psi_j] - 2{\cal D}^M F_{MN} \psi^j\gamma^N\psi_j \bigg\}.
\end{eqnarray}

The hypermultiplet part of the action includes the covariant-box operator $\sB$ given by Eq. (\ref{Box}).
In particular, it includes the derivative $\nabla^{--}$, which, in turn, contains the superfield $V^{--}$.
That is why for calculating the hypermultiplet action we need the whole component expansion of $V^{--}$.
The calculation is rather tedious, though straightforward. The linear terms are found just by making the harmonic integration in the first term (with $n=1$) of Eq. (\ref{V--_Definition}).
Next, the result can be rewritten as a function of the analytic coordinates $x^M_{\cal A}$. After that, the usual derivatives of $\psi$ and $d$
are to be replaced with the covariant ones. For the gauge field the usual derivatives are replaced with the gauge field strength.
Finally, we calculated the remaining nonlinear terms by solving the condition (\ref{Flatness}). The final result has the form

\begin{eqnarray}\label{V--_Components}
&&\hspace*{-6mm} V^{--} = \frac{1}{2} \theta^- \gamma^M \theta^- A_M + \frac{1}{\sqrt{2}} \varepsilon_{abcd}
\theta^{-a} \theta^{-b} \theta^{+c} \psi^{-d} - \sqrt{2} (\theta^-)_a^3 \psi^{+a} + \frac{i}{2} (\theta^-)^3_a (\theta^+\gamma^M\widetilde\gamma^N)^a F_{MN}\nonumber\\
&&\hspace*{-6mm}\qquad - \frac{i}{4}\varepsilon_{abcd}\theta^{-a} \theta^{-b} \theta^{+c} \theta^{+d} d^{--} +i (\theta^-)^3_a \theta^{+a} d^{+-} +i (\theta^-)^4 d^{++}
+ \frac{i}{\sqrt{2}} (\theta^-)^3_a \theta^+\gamma^M\theta^+ {\cal D}_M \psi^{-a}\nonumber\\
&&\hspace*{-6mm}\qquad + i\sqrt{2} (\theta^-)^3_a \theta^{+a}\, \theta^+\gamma^M {\cal D}_M\psi^- + i\sqrt{2} (\theta^-)^4 \theta^+\gamma^M {\cal D}_M \psi^+ + (\theta^-)^3_a (\theta^+)^3_b \Big((\widetilde\gamma^M)^{ab} {\cal D}_M d^{--} \vphantom{\frac{1}{2}}\nonumber\\
&&\hspace*{-6mm}\qquad + 2i \{\psi^{-a},\psi^{-b}\}\Big) + \frac{1}{2} (\theta^-)^4 \theta^+\gamma^N\theta^+ \Big({\cal D}^M F_{MN} + {\cal D}_N d^{+-}\Big)
+ i(\theta^{-})^4 \varepsilon_{abcd}\theta^{+a} \theta^{+b} \{\psi^{-c},\psi^{+d}\} \nonumber\\
&&\hspace*{-6mm}\qquad - 2\sqrt{2} (\theta^-)^3_a (\theta^+)^4 [d^{--},\psi^{-a}] + \sqrt{2} (\theta^-)^4 (\theta^+)^3_a \Big({\cal D}^M {\cal D}_M \psi^{-a} + [d^{--},\psi^{+a}]
- 2 [d^{+-},\psi^{-a}]
\vphantom{\frac{1}{2}}\nonumber\\
&&\hspace*{-6mm}\qquad
- \frac{1}{2} (\widetilde\gamma^M\gamma^N)^a{}_b [F_{MN},\psi^{-b}]\Big) + (\theta^-)^4 (\theta^+)^4\Big( i {\cal D}^M {\cal D}_M d^{--} +3i [d^{--},d^{+-}] + 2 (\gamma^M)_{ab} \nonumber\\
&&\hspace*{-6mm}\qquad
\times\{\psi^{-a},{\cal D}_M\psi^{-b}\}\Big).\vphantom{\frac{1}{2}}
\end{eqnarray}

In components, the hypermultiplet superfields can be written as

\begin{eqnarray}\label{Q+_Components}
&& q^{+} = q_0^{+} + \theta^{+a} q_{1a} + \theta^+ \gamma^M\theta^+ q_{2M}^{-} + (\theta^+)^3_a q_3^{(-2)a} + (\theta^+)^4 q_4^{(-3)},\nonumber\\
&& \widetilde q^{+} = \bar{q}_0^{+} + \theta^{+a} \bar{q}_{1a} + \theta^+ \gamma^M\theta^+ \bar{q}_{2M}^{-} + (\theta^+)^3_a \bar{q}_3^{(-2)a} + (\theta^+)^4 \bar{q}_4^{(-3)},
\end{eqnarray}

\noindent
where all components depend on $x_{\cal A}^M$ and harmonics. According to \cite{Ivanov:2005kz},
in the case of the higher derivative hypermultiplet the infinite number of fields in the harmonic expansions of the components  becomes dynamical.
That is why the component form of the action includes the integral over $du$. However, the integration over $d^4\theta^+$ can still be performed.

As the starting point, we consider the covariant derivative $\nabla^{++} q^+$,

\begin{equation}\label{D_Analytic}
D^{++} = \partial^{++} + \frac{i}{2} \theta^+ \gamma^M \theta^+ + \theta^{+a} \partial_{-a},\qquad D^{--} = \partial^{--} + \frac{i}{2}\theta^-\gamma^M\theta^- \partial_M + \theta^{-a} \partial_{+a},
\end{equation}

\noindent
where $\partial^{++}$ and $\partial^{--}$ act only on the explicit harmonics. The derivatives $\partial_{\pm a}$ act only on the explicit $\theta$-s according to the rules
$\partial_{+a} \theta^{+b}=\delta_a^b$, $\partial_{+a} \theta^{-b}= 0$, $\partial_{-a} \theta^{-b}=\delta_a^b$, and $\partial_{-a} \theta^{+b}=0$.
In accord with Eqs. (\ref{V++_Components}) and (\ref{D_Analytic}), the component form of the derivative $\nabla^{++} q^+$ can be obtained by substituting Eq. (\ref{Q+_Components})
into the expression

\begin{equation}
\nabla^{++}q^+ = \partial^{++} q^+ + \frac{i}{2} \theta^+\gamma^M\theta^+ {\cal D}_M q^+ + 2i\sqrt{2} (\theta^+)^3_a q^+ - 3 (\theta^+)^4 [d^{--},q^+].
\end{equation}

\noindent
Next, it is necessary to apply the operator $\sB$ to the resulting expression. This can be made using the identities like

\begin{eqnarray}\label{Box_Identities}
&&\hspace*{-5mm} \sB\Big[(\theta^+)^4 \varphi(x_{\cal A},u)\Big] = i\sqrt{2} (\theta^+)^3_a \psi^{+a} \varphi + (\theta^+)^4 \Big({\cal D}^M {\cal D}_M \varphi - d^{++} \partial^{--}\varphi - 3 d^{+-}\varphi\Big),
\nonumber\\
&&\hspace*{-5mm} \sB\Big[(\theta^+)_a^3 \varphi(x_{\cal A},u)\Big] = \frac{i}{\sqrt{2}}\varepsilon_{abcd} \theta^{+c}\theta^{+d} \psi^{+b}\varphi
+ (\theta^+)^3_a\Big({\cal D}^M {\cal D}_M\varphi - d^{++}\partial^{--}\varphi - 2 d^{+-}\varphi\Big) \nonumber\\
&&\hspace*{-5mm} + \frac{1}{2} (\theta^+)^3_b (\widetilde\gamma^N\gamma^M)^b{}_a F_{MN}\varphi
+ \sqrt{2} (\theta^+)^4\Big((\gamma^M {\cal D}_M\psi^+)_a \partial^{--}\varphi + (\gamma^M {\cal D}_M\psi^-)_a\varphi - (\gamma^M\psi^-)_a {\cal D}_M\varphi\Big),\nonumber\\
\end{eqnarray}

\noindent
etc., which can be derived with the help of Eqs. (\ref{Box}) and (\ref{V--_Components}). (Three other identities are much larger and are not presented here.) In Eq. (\ref{Box_Identities}) $\varphi$ is a (harmonic dependent) field in an arbitrary representation of the gauge group. In this case its covariant derivative is given by ${\cal D}_M\varphi = \partial_M\varphi + A_M \varphi$. For the adjoint representation the products should be replaced by commutators (or by anticommutators for products of $\psi$ with Grassmann-odd $\varphi$).

After passing through these steps, the component form of the action (\ref{Action}) can be straightforwardly deduced. It should be pointed out  that
(according to ref. \cite{Ivanov:2005kz}) the higher-derivative hypermultiplet term contains an infinite number of fields. These fields are coefficients at higher harmonic monomials
and they become dynamical in the theory with higher derivatives. This is why the harmonic integral should be left in the component action. The result includes the covariant
extension of the expression for the free action presented in \cite{Ivanov:2005kz} and a huge number of terms containing various (anti)commutators. Unfortunately,
the whole expression for the higher-derivative hypermultiplet action is too large to be presented here\footnote{In contrast to its concise superfield counterpart.}.
That is why we explicitly write down only the component form of the last term containing $F^{++}$ in Eq. (\ref{Action}) (for $\xi=1$),

\begin{eqnarray}\label{Component_Form}
&& -\frac{2i}{e_0^2}\, \mbox{tr}\int d\zeta^{(-4)} \widetilde q^{+} \big[F^{++},q^+\big] \to -\frac{2i}{e_0^2}\, \mbox{tr} \int d^6x\,du\,\bigg\{
i \bar q_4^{(-3)} \big[d^{++},q_0^+\big] + i \bar q_0^+ \big[d^{++},q_4^{(-3)}\big]\nonumber\\
&& + i \bar q_3^{(-2)a}\big[d^{++},q_{1a}\big] -i \bar q_{1a} \big[d^{++},q_3^{(-2)a}\big] - 8i \bar q_2^{-M} \big[d^{++},q_{2M}^-\big]
+i \bar q_0^+ \big[{\cal D}_M {\cal D}^M d^{--},q_0^+\big]\vphantom{\frac{1}{2}}\nonumber\\
&& + 3i\bar q_0^+ \big[\big[d^{--}, d^{+-}\big],q_0^+\big] + 2\bar q_0^+ (\gamma^M)_{ab}\big[\big\{\psi^{-a},{\cal D}_M\psi^{-b}\big\},q_0^+\big]
+ 4 \bar q_2^{-M}\big[{\cal D}^L F_{LM} + {\cal D}_M d^{+-},q_0^+\big] \vphantom{\frac{1}{2}}\nonumber\\
&& + 4 \bar q_0^+\big[{\cal D}_L F^{LM} + {\cal D}^M d^{+-},q_{2M}^-\big] + \bar q_{1a} (\widetilde\gamma^M)^{ab} \big[{\cal D}^L F_{LM} + {\cal D}_M d^{+-},q_{1b}\big]
\vphantom{\frac{1}{2}}\nonumber\\
&& + \frac{1}{\sqrt{2}} \bar q_{1a} (\widetilde\gamma^K\gamma^M)^a{}_b \big[\big[F_{KM},\psi^{-b}\big],q_0^+\big]
- \frac{1}{\sqrt{2}} \bar q_0^+ (\widetilde\gamma^K\gamma^M)^a{}_b \big\{\big[F_{KM},\psi^{-b}\big],q_{1a}\big\}
\nonumber\\
&& -\sqrt{2}\, \bar q_{1a}\big[{\cal D}_M {\cal D}^M\psi^{-a},q_0^+\big]
+ \sqrt{2}\, \bar q_0^+ \big\{{\cal D}_M {\cal D}^M\psi^{-a},q_{1a}\big\}
 -\sqrt{2}\,\bar q_{1a}\big[\big[d^{--},\psi^{+a}\big],q_0^+\big]
\vphantom{\frac{1}{2}}\nonumber\\
&&
+ \sqrt{2}\,\bar q_0^+ \big\{\big[d^{--},\psi^{+a}\big],q_{1a}\big\}
+2\sqrt{2}\,\bar q_{1a}\big[\big[d^{+-},\psi^{-a}\big],q_0^+\big]
-2\sqrt{2}\,\bar q_0^+ \big\{\big[d^{+-},\psi^{-a}\big],q_{1a}\big\}
\vphantom{\frac{1}{2}}\nonumber\\
&&
+ i\sqrt{2}\,\bar q_3^{(-2)a} (\gamma^M)_{ab} \big[{\cal D}_M\psi^{+b},q_0^+\big]
- i\sqrt{2}\,\bar q_0^+ (\gamma^M)_{ab}\big\{{\cal D}_M\psi^{+b},q_3^{(-2)a}\big\}
\vphantom{\frac{1}{2}}\nonumber\\
&& -2i\sqrt{2}\,\bar q_{2N}^- (\widetilde\gamma^N \gamma^M)^b{}_a \big\{{\cal D}_M\psi^{+a}, q_{1b}\big\}
+2i\sqrt{2}\,\bar q_{1b} (\widetilde\gamma^N \gamma^M)^b{}_a \big[{\cal D}_M\psi^{+a}, q_{2N}^-\big]
\vphantom{\frac{1}{2}}\nonumber\\
&& +4i \bar q_{2M}^- (\gamma^M)_{ab} \big[\big\{\psi^{-a},\psi^{+b}\big\},q_0^+\big]
-4i \bar q_0^+ (\gamma^M)_{ab} \big[\big\{\psi^{-a},\psi^{+b}\big\},q_{2M}^-\big]
\vphantom{\frac{1}{2}}\nonumber\\
&& -2i \bar q_{1a} \big[\big\{\psi^{-b},\psi^{+a}\big\},q_{1b}\big]
+2i \bar q_{1a} \big[\big\{\psi^{-a},\psi^{+b}\big\},q_{1b}\big]
\bigg\}.
\end{eqnarray}

\noindent
The terms presented in Eq. (\ref{Component_Form}) characterize the component structure of the non-minimal interaction. The full component expression (rather intricate),
can be straightforwardly calculated using the relations adduced above.

\section{Details of the supergraph calculation}
\label{Appendix_Supergraphs}
\hspace*{\parindent}

To calculate the diagrams presented in Fig. \ref{Figure_Diagrams}, we need the expressions for the hypermultiplet propagator
and vertices with two lines of the hypermultiplet and one or two lines of the background gauge superfield.

The vertices and propagators are derived directly from the quadratic part of the expression (\ref{Second_Variation}).
The term that does not depend on the background gauge superfield
determines the propagator of the quantum hypermultiplet superfield

\begin{equation}
- \frac{\delta^2 Z_0}{\delta j_1^{(+3)A} \delta \widetilde j_2^{(+3)B}}\bigg|_{j^{(+3)},\,\widetilde{j}^{(+3)}=0} = - \frac{ie_0^2}{(u^+_1u^+_2)^3}\,\delta^{AB}\,
\frac{1}{(\partial^2)^2} (D_1^+)^4 (D_2^+)^4 \delta^{14}(z_1-z_2).
\end{equation}

It is fairly straightforward to obtain vertices containing  two legs of the quantum hypermultiplet superfield and one leg of the background gauge superfield.
It is convenient to present this vertex as a sum of two terms (denoted by $\mbox{Ver}_1$ and $\mbox{Ver}_2$) of different origin. Namely, the expression

\begin{eqnarray}\label{Triple_Vertex_1}
&& \mbox{Ver}_1 = \frac{1}{e_0^2} f^{ABC} \int d^{(-4)}\zeta\, \widetilde{q}^{+A} \partial^2 \left(\bm{V}^{++B} q^{+C}\right) \nonumber\\
&&  + \frac{1}{2e_0^2} f^{ABC} \int d^{14}z\, du\, \widetilde{q}^{+A}\Big[2 \bm{V}_{\mbox{\scriptsize linear}}^{--B}D^{--}D^{++}q^{+C}+(D^{--}\bm{V}_{\mbox{\scriptsize linear}}^{--B})D^{++}q^{+C} \Big],\quad
\end{eqnarray}

\noindent
where $\bm{V}^{--}_{\mbox{\scriptsize linear}}$ is a part of $\bm{V}^{--}$ linear in $\bm{V}^{++}$, comes from the first term in the hypermultiplet action
in the total action (\ref{Action}). The second part of the vertex,

\begin{eqnarray}\label{Triple_Vertex_2}
&& \mbox{Ver}_2 = \frac{{\xi}}{e_0^2} f^{ABC} \int d^{(-4)}\zeta\, \widetilde{q}^{+A} \bm{F}^{++B}_{\mbox{\scriptsize linear}}q^{+C}, \quad
\end{eqnarray}

\noindent
arises due to the additional non-minimal interaction term in the hypermultiplet action. The vertex quadratic in the background gauge superfield
which comes from the expression (\ref{Second_Variation}) reads

\begin{eqnarray}\label{Vertex_3}
&& \mbox{Ver}_3 = - \frac{1}{2e_0^2} f^{ABC}f^{CDE} \int d^{14}z\, du\, \, \widetilde{q}^{+A}\Big[\, \bm{V}^{--B}_{\mbox{\scriptsize linear}}\bm{V}^{--D}_{\mbox{\scriptsize linear}}D^{++}q^{+E} +(D^{--}\bm{V}^{--B}_{\mbox{\scriptsize linear}}) \nonumber \\
&& \times \bm{V}^{++D}q^{+E} +2\bm{V}^{--B}_{\mbox{\scriptsize linear}} D^{--}\big(\bm{V}^{++D}q^{+E}\big)\Big] + \frac{\xi}{e_0^2} f^{ABC} \int d^{-4}\zeta\, \widetilde{q}^{+A} \bm{F}^{++B}_{\mbox{\scriptsize quadratic}}q^{+C}\nonumber\\
&& + \frac{1}{2e_0^2} f^{ABC} \int d^{14}z\, du\, \widetilde{q}^{+A}\Big[2 \bm{V}_{\mbox{\scriptsize quadratic}}^{--B}D^{--}D^{++}q^{+C}+(D^{--}\bm{V}_{\mbox{\scriptsize quadratic}}^{--B})D^{++}q^{+C} \Big].\qquad
\end{eqnarray}

Let us first consider the diagram (1) in Fig. \ref{Figure_Diagrams}. Depending on which vertices are present on the external lines,
it splits into the three subgraphs: $\mbox{diagram} (1) [\mbox{Ver}_1, \mbox{Ver}_1]$, $\mbox{diagram} (1) [\mbox{Ver}_1, \mbox{Ver}_2]$, and $\mbox{diagram} (1) [\mbox{Ver}_2, \mbox{Ver}_2]$.
Taking into account that  any perturbative contribution to the effective action can be written as an integral over $d^{14}z$,
the terms quadratic in $\bm{F}^{++}$ correspond to the convergent loop integrals and do not affect the divergent part of the effective action.
This implies that the contribution of $\mbox{diagram} (1) [\mbox{Ver}_2, \mbox{Ver}_2]$ is UV finite and so can be omitted. Therefore, non-trivial divergent contributions
to the effective action can appear only from the (quadratically divergent) subgraph $\mbox{diagram} (1) [\mbox{Ver}_1, \mbox{Ver}_1]$ and (logarithmically divergent)
subgraph $\mbox{diagram} (1) [\mbox{Ver}_1, \mbox{Ver}_2]$.

For the sake of convenience, we start with the calculation of the contribution of $\mbox{diagram} (1) [\mbox{Ver}_1, \mbox{Ver}_2]$. Applying Feynman rules, we find that it is given by the expression

\begin{eqnarray}\label{Nontrivial_Superdiagram2}
&&\hspace*{-5mm} \mbox{diagram} (1) [\mbox{Ver}_1, \mbox{Ver}_2] =  i {\xi} C_2 \int d^{14}z_1 \, d^{14}z_2 \, du_1\, du_2\,  \nonumber\\
&&\hspace*{-5mm} \times\frac{\bm{V}^{++A}(z_1,u_1)\,\bm{F}_{\mbox{\scriptsize linear}}^{++A}(z_2,u_2)}{(u_1^+ u_2^+)^3(u_2^+ u_1^+)^3}\,\Big[\frac{1}{\partial^2}\,\delta^{14}(z_1-z_2)\,\Big]\Big[\frac{1}{(\partial^2)^2}(D^+_2)^4(D^+_1)^4\,\delta^{14}(z_2-z_1)\,\Big]
\nonumber\\
&&\hspace*{-5mm} + i {\xi} C_2 \int d^{14}z_1 \, d^{14} z_2 \, du_1\, du_2\,\bm{V}_{\mbox{\scriptsize linear}}^{--A}(z_1,u_1)\,\bm{F}_{\mbox{\scriptsize linear}}^{++A}(z_2,u_2)  \nonumber\\
&&\hspace*{-5mm} \times \, D_1^{--}D_1^{++}\Big[\frac{1}{(u_1^+ u_2^+)^3} \frac{1}{(\partial^2)^2}(D_1^{+})^4(D_2^{+})^4\delta^{14}(z_1-z_2)\,\Big]\,\Big[ \frac{1}{(u_2^+ u_1^+)^3}\frac{1}{(\partial^2)^2}(D_2^{+})^4\delta^{14}(z_1-z_2)\,\Big]
\nonumber\\
&&\hspace*{-5mm} + \frac{i {\xi} C_2}{2} \int d^{14}z_1 \, d^{14} z_2 \, du_1\, du_2\,D_1^{--}\bm{V}_{\mbox{\scriptsize linear}}^{--A}(z_1,u_1)\,\bm{F}_{\mbox{\scriptsize linear}}^{++A}(z_2,u_2)  \nonumber\\
&&\hspace*{-5mm} \times \, D_1^{++}\Big[\frac{1}{(u_1^+ u_2^+)^3} \frac{1}{(\partial^2)^2}(D_1^{+})^4(D_2^{+})^4\delta^{14}(z_1-z_2)\,\Big]\,\Big[ \frac{1}{(u_2^+ u_1^+)^3}\frac{1}{(\partial^2)^2}(D_2^{+})^4\delta^{14}(z_1-z_2)\,\Big],
\qquad
\end{eqnarray}

\noindent
which is linear in ${\xi}$. In the second and third terms of this expression we used the identity

\begin{equation}
D^{++}_1\frac{1}{(u^+_1u^+_2)^n} = \frac{1}{(n-1)!}(D^{--}_1)^{n-1}\delta^{(n,-n)}(u_1,u_2)
\end{equation}

\noindent
and took into account that

\begin{equation}\label{identity}
\frac{1}{2}\delta^{(3,-3)}(u_1,u_2)(D_1^{--})^2(D_1^{+})^4(D_2^{+})^4 \delta^{14}(z_1-z_2)  = \delta^{(3,-3)}(u_1,u_2) (D_2^{+})^4 \partial^2 \delta^{14}(z_1-z_2).
\end{equation}

Without loss of generality, we can assume that all spinor derivatives act on the point $z_1$. The $\delta$-functions $\delta^{14}(z_1-z_2)$ include $\delta^8(\theta_1-\theta_2)$, and the product of two Grassmannian
$\delta$-functions does not vanish only in the situation when at least 8 spinor covariant derivatives act on them. Taking this into account and using the identity (\ref{formula}) we can do one of the integrals over $d^8\theta$. Then, after calculating the harmonic integrals in the momentum representation, the expression (\ref{Nontrivial_Superdiagram2}) takes the form

\begin{eqnarray}
&& \mbox{diagram} (1) [\mbox{Ver}_1, \mbox{Ver}_2]\nonumber\\
&&\qquad =  2i \xi C_2 \, \int \frac{d^6p}{(2\pi)^6}\, d^8\theta\, du\,\bm{V}_{\mbox{\scriptsize linear}}^{--A}(-p,\theta,u)\, \bm{F}_{\mbox{\scriptsize linear}}^{++A}(p,\theta,u) \int \frac{d^6k}{(2\pi)^6}\,\frac{1}{k^4 (k+p)^2}.\qquad
\end{eqnarray}

The momentum integral here is calculated in the Euclidean space after the Wick rotation with the help of dimensional reduction. Clearly, we are interested in its divergent part only,

\begin{equation}
\int \frac{d^6k}{(2\pi)^6} \frac{1}{k^4 (k+p)^2}\ \to\ -i\int \frac{d^DK}{(2\pi)^D} \frac{1}{K^6} + \mbox{finite terms} = -\frac{i}{\varepsilon (4\pi)^3}  + \mbox{finite terms},
\end{equation}

\noindent
where the capital letter $K$ denotes the Euclidean loop momentum. Therefore, in the coordinate representation the divergent contribution coming from the $\mbox{diagram} (1) [\mbox{Ver}_1, \mbox{Ver}_2]$ part of the supergraph (1) in Fig. \ref{Figure_Diagrams} is given by the expression

\begin{equation}\label{One_Loop_Divergence_Quadratic}
\frac{4\xi C_2}{\varepsilon(4\pi)^3} \mbox{tr} \int d\zeta^{(-4)}\, \big(\bm{F}_{\mbox{\scriptsize linear}}^{++}\big)^2.
\end{equation}

Now, let us calculate the subgraph $\mbox{diagram} (1) [\mbox{Ver}_1,\mbox{Ver}_1]$. Using the Feynman rules we present it in the form

\begin{eqnarray}\label{Nontrivial_Superdiagram1}
&&\hspace*{-7mm} \mbox{diagram} (1) [\mbox{Ver}_1, \mbox{Ver}_1]\nonumber\\
&&\hspace*{-7mm} =  \frac{i C_2}{2} \int d^{14}z_1 \, d^{14}z_2 \, du_1\, du_2\,\bigg\{ \frac{\bm{V}^{++A}(z_1,u_1)\,\bm{V}^{++A}(z_2,u_2)}{(u_1^+ u_2^+)^3(u_2^+ u_1^+)^3} \, \Big[\frac{1}{\partial^2}(D^+_1)^4(D^+_2)^4\,\delta^{14}(z_1-z_2)\,\Big] \nonumber\\
&&\hspace*{-7mm} \times \Big[\frac{1}{\partial^2}\,\delta^{14}(z_1-z_2)\,\Big]
+ \bm{V}^{++A}(z_1,u_1)\,\Big[ \frac{1}{(u_1^+u_2^+)^3} \frac{1}{(\partial^2)^2}(D_2^{+})^4\delta^{14}(z_1-z_2)\,\Big]\, \Big[2 \bm{V}_{\mbox{\scriptsize linear}}^{--A}(z_2,u_2)  D_2^{--} \nonumber\\
&&\hspace*{-7mm}  + (D_2^{--}\bm{V}_{\mbox{\scriptsize linear}}^{--A}(z_2,u_2))\Big]
D_2^{++}\Big[\frac{1}{(u_2^+ u_1^+)^3} \frac{1}{\partial^2}(D_1^{+})^4(D_2^{+})^4\delta^{14}(z_1-z_2)\,\Big] + \frac{1}{4} \Big[2 \bm{V}_{\mbox{\scriptsize linear}}^{--A}(z_1,u_1)  D_1^{--}\nonumber\\
&&\hspace*{-7mm}  + (D_1^{--}\bm{V}_{\mbox{\scriptsize linear}}^{--A}(z_1,u_1))\Big]
D_1^{++}\Big[\frac{1}{(u_1^+ u_2^+)^3} \frac{1}{(\partial^2)^2}(D_1^{+})^4(D_2^{+})^4\delta^{14}(z_1-z_2)\,\Big]\, \Big[2 \bm{V}_{\mbox{\scriptsize linear}}^{--A}(z_2,u_2)  D_2^{--} \nonumber\\
&&\hspace*{-7mm} + (D_2^{--}\bm{V}_{\mbox{\scriptsize linear}}^{--A}(z_2,u_2))\Big]
D_2^{++}\Big[\frac{1}{(u_2^+ u_1^+)^3} \frac{1}{(\partial^2)^2}(D_1^{+})^4(D_2^{+})^4\delta^{14}(z_1-z_2)\,\Big] \bigg\}.
\end{eqnarray}

\noindent
After applying the identities (\ref{formula}), (\ref{identity}) and calculating one of the integrals over $d^8\theta$,
in the momentum representation the expression (\ref{Nontrivial_Superdiagram1}) can be cast in the form

\begin{eqnarray}\label{quadratic_divergence}
&&\hspace*{-5mm} \mbox{diagram} (1) [\mbox{Ver}_1, \mbox{Ver}_1]\nonumber\\
&&\hspace*{-5mm} =  -\frac{i C_2}{2}\, \int \frac{d^6p}{(2\pi)^6}\, d^8\theta \, \frac{du_1\, du_2}{(u^+_1 u_2^+)^2}\, \bm{V}^{++A}(-p,\theta ,u_1) \bm{V}^{++A}(p,\theta ,u_2)
\int \frac{d^6k}{(2\pi)^6}\bigg(\frac{1}{k^2 (k+p)^2}+\frac{2}{k^4}\bigg).\nonumber\\
\end{eqnarray}

Next, we proceed to the calculation of the (quadratically divergent) tadpole contribution corresponding to the diagram (2) in Fig.~\ref{Figure_Diagrams}. The vertex $\mbox{Ver}_3$ in this diagram is given by Eq. (\ref{Vertex_3}). Employing the identities (\ref{formula}) and (\ref{identity}) for the evaluation, after some transformations in the momentum representation we obtain the expression

\begin{equation}
i C_2 \, \int \frac{d^6p}{(2\pi)^6}\, d^8\theta \, \frac{du_1\, du_2}{(u^+_1 u_2^+)^2} \, \bm{V}^{++A}(-p,\theta ,u_1) \bm{V}^{++A}(p,\theta ,u_2) \int \frac{d^6k}{(2\pi)^6}\frac{1}{k^4},
\end{equation}

\noindent
which cancels the second term in (\ref{quadratic_divergence}), so that

\begin{eqnarray}\label{Quadratic}
&&\hspace*{-5mm} \mbox{diagram} (1) [\mbox{Ver}_1, \mbox{Ver}_1] + \mbox{diagram} (2)\nonumber\\
&&\hspace*{-5mm} =  -\frac{i C_2}{2}\, \int \frac{d^6p}{(2\pi)^6}\, d^8\theta \, \frac{du_1\, du_2}{(u^+_1 u_2^+)^2}\, \bm{V}^{++A}(-p,\theta ,u_1) \bm{V}^{++A}(p,\theta ,u_2)
\int \frac{d^6k}{(2\pi)^6}\,\frac{1}{k^2 (k+p)^2}.\qquad
\end{eqnarray}

\noindent
This expression exactly cancels the contribution of the Faddeev--Popov and Nielsen--Kallosh ghosts (see, e.g., \cite{Buchbinder:2017ozh}), which is also quadratically divergent.
All other contributions diverge only logarithmically. Therefore, there are no quadratic divergences in the theory under consideration, at least, in the one-loop approximation, for any value of ${\xi}$.

When using the regularization by dimensional reduction, applying the standard Euclidean technique, we obtain

\begin{equation}\label{Integral_Divergent_Part}
\int \frac{d^6k}{(2\pi)^6} \frac{1}{k^2 (k+p)^2} \to i \int \frac{d^DK}{(2\pi)^D} \frac{1}{K^2 (K+P)^2} = -\frac{iP^2}{3\varepsilon (4\pi)^3} + \mbox{finite terms}.
\end{equation}

\noindent
Taking into account that $P^2 = - p^2$, the contribution to the divergent part of the effective action under consideration can be written as

\begin{equation}\label{Logarithmic_Divergence}
C_2\,\int \frac{d^6p}{(2\pi)^6}\, d^8\theta\, \frac{du_1\, du_2}{(u_1^+ u_2^+)^2}\, \bm{V}^{++A}(-p,\theta,u_1) \bm{V}^{++A}(p,\theta,u_2)\, \frac{p^2}{6\varepsilon (4\pi)^3}.
\end{equation}

\noindent
There is only one way to bring this expression to a gauge-invariant form,

\begin{eqnarray}\label{Invariant}
&& \mbox{tr}\int d\zeta^{(-4)}\, (\bm{F}^{++})^2 = \mbox{tr} \int d^{14}z\,du\, \bm{V}^{--} \partial^2 \bm{V}^{++} + O\left((\bm{V}^{++})^3\right)\nonumber\\
&&\qquad\qquad\qquad = \mbox{tr} \int d^{14}z\,\frac{du_1\,du_2}{(u_1^+ u_2^+)^2}\, \bm{V}^{++}(z,u_1)\partial^2 \bm{V}^{++}(z,u_2) + O\left((\bm{V}^{++})^3\right).\qquad
\end{eqnarray}

\noindent
This implies that the manifestly background gauge-invariant generalization of the expression (\ref{Logarithmic_Divergence}) is written as

\begin{equation}\label{Simple_Hypermultiplet}
-\frac{C_2}{3\varepsilon (4\pi)^3}\, \mbox{tr} \int d\zeta^{(-4)}\, (\bm{F}^{++})^2.
\end{equation}

\noindent
The sum of the expressions (\ref{One_Loop_Divergence_Quadratic}) and (\ref{Simple_Hypermultiplet})
gives the total hypermultiplet contribution (\ref{Divergent_Contribution}) for the model (\ref{Action}).


\begin{thebibliography}{150}

\bibitem{Howe:1983jm}
P.S.~Howe and K.S.~Stelle,
``Ultraviolet Divergences in Higher Dimensional Supersymmetric {Yang-Mills} Theories,''
Phys.\ Lett.\ B {\bf 137} (1984), 175.

\bibitem{Howe:2002ui}
P.S.~Howe and K.S.~Stelle,
``Supersymmetry counterterms revisited,''
Phys.\ Lett.\ B {\bf 554} (2003), 190
 [arXiv:hep-th/0211279 [hep-th]].

\bibitem{Bossard:2009sy}
G.~Bossard, P.S.~Howe and K.S.~Stelle,
``The Ultra-violet question in maximally supersymmetric field theories,''
Gen.\ Rel.\ Grav.\ {\bf 41} (2009), 919
 [arXiv:0901.4661 [hep-th]].

\bibitem{Bossard:2009mn}
G.~Bossard, P.S.~Howe and K.S.~Stelle,
``A Note on the UV behaviour of maximally supersymmetric Yang-Mills theories,''
Phys.\ Lett.\ B {\bf 682} (2009), 137
 [arXiv:0908.3883 [hep-th]].

\bibitem{Fradkin:1982kf}
E.S.~Fradkin and A.A.~Tseytlin,
``Quantum Properties of Higher Dimensional and Dimensionally Reduced Supersymmetric Theories,''
Nucl.\ Phys.\ B {\bf 227} (1983), 252.

\bibitem{Marcus:1983bd}
N.~Marcus and A.~Sagnotti,
``A Test of Finiteness Predictions for Supersymmetric Theories,''
Phys.\ Lett.\ B {\bf 135} (1984), 85.

\bibitem{Smilga:2016dpe}
A.~Smilga,
``Ultraviolet divergences in non-renormalizable supersymmetric theories,''
Phys.\ Part.\ Nucl.\ Lett.\ {\bf 14} (2017) no.2, 245
 [arXiv:1603.06811 [hep-th]].

\bibitem{Bork:2015zaa}
L.V.~Bork, D.I.~Kazakov, M.V.~Kompaniets, D.M.~Tolkachev and D.E.~Vlasenko,
``Divergences in maximal supersymmetric Yang-Mills theories in diverse dimensions,''
JHEP {\bf 11} (2015), 059
 [arXiv:1508.05570 [hep-th]].

\bibitem{Pais:1950za}
A.~Pais and G.E.~Uhlenbeck,
``On Field theories with nonlocalized action,''
Phys.\ Rev.\ {\bf 79} (1950), 145.

\bibitem{Stelle:1976gc}
K.S.~Stelle,
``Renormalization of Higher Derivative Quantum Gravity,''
Phys.\ Rev.\ D {\bf 16} (1977), 953.

\bibitem{Galperin:1984av}
A.~Galperin, E.~Ivanov, S.~Kalitzin, V.~Ogievetsky and E.~Sokatchev,
``Unconstrained N=2 Matter, Yang-Mills and Supergravity Theories in Harmonic Superspace,''
Class. Quant. Grav. \textbf{1} (1984), 469\ 
[erratum: Class. Quant. Grav. \textbf{2} (1985), 127].

\bibitem{Galperin:1985bj}
A.~Galperin, E.A.~Ivanov, V.~Ogievetsky and E.~Sokatchev,
``Harmonic Supergraphs. Green Functions,''
Class. Quant. Grav. \textbf{2} (1985), 601.

\bibitem{Galperin:1985va}
A.~Galperin, E.~Ivanov, V.~Ogievetsky and E.~Sokatchev,
``Harmonic Supergraphs. Feynman Rules and Examples,''
Class. Quant. Grav. \textbf{2} (1985), 617.

\bibitem{Galperin:2001uw}
A.S.~Galperin, E.A.~Ivanov, V.I.~Ogievetsky and E.S.~Sokatchev,
``Harmonic superspace'',
Cambridge, UK: Univ. Pr. (2001) 306 p.

\bibitem{Ivanov:2005qf}
E.A.~Ivanov, A.V.~Smilga and B.M.~Zupnik,
``Renormalizable supersymmetric gauge theory in six dimensions,''
Nucl. Phys. B \textbf{726} (2005), 131\ 
 [arXiv:hep-th/0505082 [hep-th]].

\bibitem{Casarin:2019aqw}
L.~Casarin and A.A.~Tseytlin,
``One-loop $\beta$-functions in 4-derivative gauge theory in 6 dimensions,''
JHEP \textbf{08} (2019), 159
 [arXiv:1907.02501 [hep-th]].

\bibitem{Buchbinder:2020tnc}
I.L.~Buchbinder, E.A.~Ivanov, B.S.~Merzlikin and K.V.~Stepanyantz,
``Supergraph calculation of one-loop divergences in higher-derivative $6D$ SYM theory,''
JHEP \textbf{08} (2020), 169
 [arXiv:2004.12657 [hep-th]].

\bibitem{Buchbinder:2020ovf}
I.L.~Buchbinder, E.A.~Ivanov, B.S.~Merzlikin and K.V.~Stepanyantz,
``The renormalization structure of $6D$, ${\cal N}=(1,0)$ supersymmetric higher-derivative gauge theory,''
Nucl. Phys. B \textbf{961} (2020), 115249
 [arXiv:2007.02843 [hep-th]].

\bibitem{Buchbinder:2016gmc}
I.L.~Buchbinder, E.A.~Ivanov, B.S.~Merzlikin and K.V.~Stepanyantz,
``One-loop divergences in the $6D, \mathcal N = (1,0)$ abelian gauge theory,''
Phys. Lett. B \textbf{763} (2016), 375
 [arXiv:1609.00975 [hep-th]].

\bibitem{Buchbinder:2016url}
I.L.~Buchbinder, E.A.~Ivanov, B.S.~Merzlikin and K.V.~Stepanyantz,
``One-loop divergences in 6D, $ \mathcal{N} $ = (1, 0) SYM theory,''
JHEP \textbf{01} (2017), 128
 [arXiv:1612.03190 [hep-th]].

\bibitem{Buchbinder:2017ozh}
I.L.~Buchbinder, E.A.~Ivanov, B.S.~Merzlikin and K.V.~Stepanyantz,
``Supergraph analysis of the one-loop divergences in $6D$, ${\cal N} = (1,0)$ and ${\cal N} = (1,1)$ gauge theories,''
Nucl. Phys. B \textbf{921} (2017), 127\ 
 [arXiv:1704.02530 [hep-th]].

\bibitem{Budekhina:2021kbf}
A.S.~Budekhina and B.S.~Merzlikin,
``One-loop divergences in the six-dimensional N=(1,0) hypermultiplet self-coupling model,''
Phys. Rev. D \textbf{104} (2021) no.10, 106010
 [arXiv:2103.12054 [hep-th]].

\bibitem{Sohnius:1981sn}
M.F.~Sohnius and P.C.~West,
``Conformal Invariance in N=4 Supersymmetric Yang-Mills Theory,''
Phys. Lett. B \textbf{100} (1981), 245.

\bibitem{Grisaru:1982zh}
M.T.~Grisaru and W.~Siegel,
``Supergraphity. 2. Manifestly Covariant Rules and Higher Loop Finiteness,''
Nucl. Phys. B \textbf{201} (1982), 292
[erratum: Nucl. Phys. B \textbf{206} (1982), 496].

\bibitem{Howe:1983sr}
P.S.~Howe, K.S.~Stelle and P.K.~Townsend,
``Miraculous Ultraviolet Cancellations in Supersymmetry Made Manifest,''
Nucl. Phys. B \textbf{236} (1984), 125.

\bibitem{Mandelstam:1982cb}
S.~Mandelstam,
``Light Cone Superspace and the Ultraviolet Finiteness of the N=4 Model,''
Nucl. Phys. B \textbf{213} (1983), 149.

\bibitem{Brink:1982pd}
L.~Brink, O.~Lindgren and B.E.W.~Nilsson,
``N=4 Yang-Mills Theory on the Light Cone,''
Nucl. Phys. B \textbf{212} (1983), 401.

\bibitem{Zupnik:1986da}
B.M.~Zupnik,
``Six-dimensional Supergauge Theories in the Harmonic Superspace,''
Sov. J. Nucl. Phys. \textbf{44} (1986), 512
 [Yad. Fiz.  \textbf{44} (1986) 794].

\bibitem{Buchbinder:2025aac}
I.L.~Buchbinder, A.S.~Budekhina, E.A.~Ivanov and K.V.~Stepanyantz,
``The structure of divergences in the higher-derivative supersymmetric 6D gauge theory,''
Phys. Rev. D \textbf{111} (2025) no.12, 125014
 [arXiv:2503.18532 [hep-th]].

\bibitem{Kuzenko:2017xgh}
S.M.~Kuzenko, J.~Novak and I.B.~Samsonov,
``Chiral anomalies in six dimensions from harmonic superspace,''
JHEP \textbf{11} (2017), 145
 [arXiv:1708.08238 [hep-th]].

\bibitem{Smilga:2006ax}
A.V.~Smilga,
``Chiral anomalies in higher-derivative supersymmetric 6D theories,''
Phys. Lett. B \textbf{647} (2007), 298\ 
 [arXiv:hep-th/0606139 [hep-th]].

\bibitem{Novikov:1983uc}
V.A.~Novikov, M.A.~Shifman, A.I.~Vainshtein and V.I.~Zakharov,
``Exact Gell-Mann-Low Function of Supersymmetric Yang-Mills Theories from Instanton Calculus,''
Nucl. Phys. B \textbf{229} (1983), 381.

\bibitem{Jones:1983ip}
D.R.T.~Jones,
``More on the Axial Anomaly in Supersymmetric {Yang-Mills} Theory,''
Phys. Lett.  \textbf{123B} (1983), 45.

\bibitem{Novikov:1985rd}
V.A.~Novikov, M.A.~Shifman, A.I.~Vainshtein and V.I.~Zakharov,
``Beta Function in Supersymmetric Gauge Theories: Instantons Versus Traditional Approach,''
Phys. Lett.  \textbf{166B} (1986), 329
[Sov. J. Nucl. Phys.  \textbf{43} (1986), 294]
[Yad. Fiz. \textbf{43} (1986), 459].

\bibitem{Shifman:1986zi}
M.A.~Shifman and A.I.~Vainshtein,
``Solution of the Anomaly Puzzle in SUSY Gauge Theories and the Wilson Operator Expansion,''
Nucl. Phys. B \textbf{277} (1986), 456
[Sov. Phys. JETP \textbf{64} (1986), 428]
[Zh. Eksp. Teor. Fiz.  \textbf{91} (1986), 723].

\bibitem{Slavnov:1971aw}
A.A.~Slavnov,
``Invariant regularization of nonlinear chiral theories,''
Nucl. Phys. B \textbf{31} (1971), 301.

\bibitem{Slavnov:1972sq}
A.A.~Slavnov,
``Invariant regularization of gauge theories,''
Theor. Math. Phys. \textbf{13} (1972), 1064
[Teor. Mat. Fiz. \textbf{13} (1972), 174].

\bibitem{Slavnov:1977zf}
A.A.~Slavnov,
``The Pauli-Villars Regularization for Nonabelian Gauge Theories,''
Theor. Math. Phys. \textbf{33} (1977), 977
[Teor. Mat. Fiz. \textbf{33} (1977), 210].

\bibitem{Krivoshchekov:1978xg}
V.K.~Krivoshchekov,
``Invariant Regularizations for Supersymmetric Gauge Theories,''
Theor. Math. Phys. \textbf{36} (1978), 745
[Teor. Mat. Fiz.  \textbf{36} (1978), 291].

\bibitem{West:1985jx}
P.C.~West,
``Higher Derivative Regulation of Supersymmetric Theories,''
Nucl. Phys. B \textbf{268} (1986), 113.

\bibitem{Buchbinder:2015eva}
I~L.~Buchbinder, N.G.~Pletnev and K.V.~Stepanyantz,
``Manifestly N=2 supersymmetric regularization for N=2 supersymmetric field theories,''
Phys. Lett. B \textbf{751} (2015), 434\ 
 [arXiv:1509.08055 [hep-th]].

\bibitem{Stepanyantz:2019lyo}
K.~Stepanyantz,
``The Higher Covariant Derivative Regularization as a Tool for Revealing the Structure of Quantum Corrections in Supersymmetric Gauge Theories,''
Proc. Steklov Inst. Math. \textbf{309} (2020) no.1, 284\ 
 [arXiv:1910.03242 [hep-th]].

\bibitem{Stepanyantz:2023jot}
K.V.~Stepanyantz,
``The structure of quantum corrections and exact results in supersymmetric theories from the higher covariant derivative regularization,''
Theor. Math. Phys. \textbf{217} (2023) no.3, 1954.

\bibitem{Heinemeyer:2019vbc}
S.~Heinemeyer, M.~Mondragón, N.~Tracas and G.~Zoupanos,
``Reduction of Couplings and its application in Particle Physics,''
Phys. Rept.  \textbf{814} (2019), 1
 [arXiv:1904.00410 [hep-ph]].

\bibitem{Kazakov:1986bs}
D.I.~Kazakov,
``Finite $N=1$ {SUSY} Field Theories and Dimensional Regularization,''
Phys. Lett. B \textbf{179} (1986), 352.

\bibitem{Ermushev:1986cu}
A.V.~Ermushev, D.I.~Kazakov and O.V.~Tarasov,
``Finite N=1 Supersymmetric Grand Unified Theories,''
Nucl. Phys. B \textbf{281} (1987), 72.

\bibitem{Lucchesi:1987he}
C.~Lucchesi, O.~Piguet and K.~Sibold,
``Vanishing Beta Functions in $N=1$ Supersymmetric Gauge Theories,''
Helv. Phys. Acta \textbf{61} (1988), 321.

\bibitem{Lucchesi:1987ef}
C.~Lucchesi, O.~Piguet and K.~Sibold,
``Necessary and Sufficient Conditions for All Order Vanishing Beta Functions in Supersymmetric {Yang-Mills} Theories,''
Phys. Lett. B \textbf{201} (1988), 241.

\bibitem{Parkes:1984dh}
A.~Parkes and P.C.~West,
``Finiteness in Rigid Supersymmetric Theories,''
Phys. Lett. \textbf{138B} (1984), 99.

\bibitem{Jones:1983vk}
D.R.T.~Jones and L.~Mezincescu,
``The Beta Function in Supersymmetric {Yang-Mills} Theory,''
Phys. Lett. \textbf{136B} (1984), 242.

\bibitem{Jones:1984cx}
D.R.T.~Jones and L.~Mezincescu,
``The Chiral Anomaly and a Class of Two Loop Finite Supersymmetric Gauge Theories,''
Phys. Lett.  \textbf{138B} (1984), 293.

\bibitem{Grisaru:1985tc}
M.T.~Grisaru, B.~Milewski and D.~Zanon,
``The Structure of {UV} Divergences in Ssym Theories,''
Phys. Lett.  \textbf{155B} (1985), 357.

\bibitem{Stepanyantz:2021dus}
K.~Stepanyantz,
``Exact $\beta $-functions for $\mathcal{N}=1$ supersymmetric theories finite in the lowest loops,''
Eur. Phys. J. C \textbf{81} (2021) no.7, 571
 [arXiv:2105.00900 [hep-th]].

\bibitem{Ivanov:2005kz}
E.~A.~Ivanov and A.~V.~Smilga,
``Conformal properties of hypermultiplet actions in six dimensions,''
Phys. Lett. B \textbf{637} (2006), 374.\ 
 [arXiv:hep-th/0510273 [hep-th]].

\bibitem{Buchbinder:1997ya}
I.~L.~Buchbinder, E.~I.~Buchbinder, S.~M.~Kuzenko and B.~A.~Ovrut,
``The Background field method for N=2 superYang-Mills theories in harmonic superspace,''
Phys. Lett. B \textbf{417} (1998), 61.\ 
 [arXiv:hep-th/9704214 [hep-th]].

\bibitem{Buchbinder:2001wy}
E.I.~Buchbinder, B.A.~Ovrut, I.L.~Buchbinder, E.A.~Ivanov and S.M.~Kuzenko,
``Low-energy effective action in N = 2 supersymmetric field theories,''
Phys. Part. Nucl. \textbf{32} (2001), 641\ 
 [Fiz. Elem. Chast. Atom. Yadra {\bf 32} (2001) 1222].

\bibitem{Howe:1985ar}
P.~S.~Howe, K.~S.~Stelle and P.~C.~West,
``N=1 d = 6 harmonic superspace,''
Class. Quant. Grav. \textbf{2} (1985), 815.

\bibitem{Bossard:2015dva}
G.~Bossard, E.~Ivanov and A.~Smilga,
``Ultraviolet behavior of 6D supersymmetric Yang-Mills theories and harmonic superspace,''
JHEP \textbf{12} (2015), 085
 [arXiv:1509.08027 [hep-th]].

\bibitem{Buchbinder:2014sna}
I.~L.~Buchbinder and N.~G.~Pletnev,
``Construction of 6D supersymmetric field models in N=(1,0) harmonic superspace,''
Nucl. Phys. B \textbf{892} (2015), 21\ 
 [arXiv:1411.1848 [hep-th]].

\bibitem{DeWitt:1965jb}
B.S.~DeWitt\ \
``Dynamical theory of groups and fields,'' Conf. Proc. C \textbf{630701} (1964), 585
[Les Houches Lect. Notes \textbf{13} (1964), 585].

\bibitem{DeWitt:1967ub}
B.~S.~DeWitt,
``Quantum Theory of Gravity. 2. The Manifestly Covariant Theory,''
Phys. Rev. \textbf{162} (1967), 1195.

\bibitem{Kallosh:1974yh}
R.~E.~Kallosh,
``The Renormalization in Nonabelian Gauge Theories,''
Nucl. Phys. B \textbf{78} (1974), 293.

\bibitem{Arefeva:1974jv}
I.Ya.~Arefeva, L.D.~Faddeev and A.A.~Slavnov,
``Generating Functional for the s Matrix in Gauge Theories,''
Theor. Math. Phys. \textbf{21} (1974), 1165
[Teor. Mat. Fiz. \textbf{21} (1974), 311].

\bibitem{Abbott:1980hw}
L.F.~Abbott,
``The Background Field Method Beyond One Loop,''
Nucl. Phys. B \textbf{185} (1981), 189.

\bibitem{Abbott:1981ke}
L.F.~Abbott,
``Introduction to the Background Field Method,''
Acta Phys. Polon. B \textbf{13} (1982), 33
CERN-TH-3113.

\bibitem{Siegel:1979wq}
W.~Siegel,
``Supersymmetric Dimensional Regularization via Dimensional Reduction,''
Phys. Lett. B \textbf{84} (1979), 193.

\bibitem{Buyucli:2021csy}
S.~Buyucli and E.~Ivanov,
``Higher-dimensional invariants in 6D super Yang-Mills theory,''
JHEP \textbf{07} (2021), 190
 [arXiv:2105.05899 [hep-th]].

\end{thebibliography}
\end{document}